\documentclass[journal]{IEEEtran}
\usepackage[T1]{fontenc}
\usepackage{amsmath}
\usepackage{lineno,hyperref}
\usepackage{mathtools}
\usepackage{amssymb}
\usepackage{amsmath}
\newcommand\norm[1]{\left\lVert#1\right\rVert}
\usepackage{graphicx}
\usepackage{caption}
\usepackage{multirow}
\usepackage{cite}
\usepackage[caption=false]{subfig}
\hypersetup{draft}
\usepackage[usenames, dvipsnames]{color}
\DeclareMathOperator*{\argmax}{arg\,max}  
\DeclareMathOperator*{\argmin}{arg\,min}  

\usepackage[cmintegrals]{newtxmath}
\usepackage[letterpaper, left=0.625in, right=0.625in, bottom=1in, top=0.725in]{geometry}
\usepackage[vlined,linesnumbered,ruled,resetcount]{algorithm2e}
\SetKwInput{KwInput}{Input}
\SetKwInput{KwInitialization}{Initialization}
\SetKwInput{KwOutput}{Output}
\usepackage[nodisplayskipstretch]{setspace}
\newcommand{\rev}[1]{{\color{black}#1}}
\newcommand{\revv}[1]{{\color{black}#1}}
\newcommand{\revvv}[1]{{\color{black}#1}}

\begin{document}

\title{Permutation Matrix Modulation}

\author{Rahmat~Faddli~Siregar,~Nandana~Rajatheva,~and~Matti~Latva-Aho \\
	Centre for Wireless Communications, University of Oulu, Finland
\thanks{This work was supported by the Academy of Finland, 6G Flagship Program, under Grant 346208.}
\thanks{The authors are with the Centre for Wireless Communications, University of Oulu, 90570 Oulu, Finland (e-mail: rahmat.siregar@oulu.fi; nandana.rajatheva@oulu.fi; matti.latva-aho@oulu.fi).}
}

\markboth{This article has been accepted for publication in IEEE Transaction on Wireless Communications}%
{}

\maketitle

\begin{abstract}
We propose a novel scheme that allows MIMO system to modulate a set of permutation matrices to send more information bits, extending our initial work on the topic. This system is called Permutation Matrix Modulation (PMM). The basic idea is to employ a permutation matrix as a precoder and treat it as a modulated symbol. We continue the evolution of index modulation in MIMO by adopting all-antenna activation and obtaining a set of unique symbols from altering the positions of the antenna transmit power. We provide the analysis of the achievable rate of PMM under Gaussian Mixture Model (GMM) distribution \revv{and finite cardinality input (FCI). Numerical results are evaluated by comparing PMM with the other existing systems.} We also present a way to attain the optimal achievable rate of PMM by solving a maximization problem via interior-point method. A low complexity detection scheme based on zero-forcing (ZF) is proposed, and maximum likelihood (ML) detection is discussed. We demonstrate the trade-off between simulation of the symbol error rate (SER) and the computational complexity where ZF performs worse in the SER simulation but requires much less computational complexity than ML.
\end{abstract}



\section{Introduction}
\IEEEPARstart{T}{he} research on finding alternative symbols to send more information bits without utilizing expensive conventional resources (i.e., time and frequency) is a growing topic in wireless communication. Many works have been done to establish the foundation of this area known as Spatial Modulation (SM) and Generalised Spatial Modulation (GSM) in \cite{SM_achrate, SM1, SM2, GSM1} as well as its latest development called Quadrature Spatial Modulation (QSM) in \cite{QSM1,QSM3,QSM2}. The main idea of these systems is to exploit the potential \textit{"modulate-able"} symbols that are coming from the spatial characteristics of multiple input multiple output (MIMO). What makes them interesting is that they arguably can be integrated into the existing technologies (i.e., 4G and 5G) without requiring any major changes in the hardware, unlike most of the other proposed systems, which often demand a total change from the existing technologies. Furthermore, this area can also be attractive for the incoming 6G technology to be adopted.

A new system called Permutation Channel Modulation (PCM) has been proposed in \cite{PCM_faddli} which offers a new paradigm to exploit index modulation in MIMO. PCM permutes the position of singular values obtained from decomposing the MIMO channel by assuming that channel state information is available at both transmitter (CSIT) and receiver (CSIR). The position of the singular values is rearranged using a permutation matrix and treat the permutation matrix as a symbol to send more information bits. This idea is a breakthrough since the number of \rev{maximum transmit bits}\footnote{\rev{Let us clarify that maximum transmit bits is the maximum number of bits a transmitter can radiate per transmission. It is obviously not the same with the terms "throughput" as \cite{5684112} and \cite{5688440} used. It is also not "spectral efficiency" as \cite{6879496} used. We understand that the general definitions of data rate, throughput and spectral efficiency are the performance metrics that are measured by defining the mutual information of the transmit signal and the receive signal. To be more clear, when we employ 64-QAM, it does not mean we have 6 bits/s throughput or 6 bits/Hz spectral efficiency. It just means that one constellation point represents 6 bits, and thus, we can transmit 6 bits for every symbol we radiate. We believe clarifying this issue will avoid confusions to the readers.}} substantially increases compared to the previous systems (i.e., SM, GSM, and QSM). It can be seen in Fig. \ref{extra_bits} where we plot the number of maximum transmit bits versus the number of transmit antennas. \rev{Fig. \ref{extra_bits} is computed based on the maximum transmit bits of each system provided in Table \ref{max_txbits}.} Note that Fig. \ref{extra_bits} does not necessarily mean the rate, but it can be interpreted as the potential rate of each corresponding system in providing the \textit{"modulate-able"} symbols/indices.

\subsection{Motivation and Previous Works}
There are two main benefits when employing SM over MIMO, as mentioned in \cite{SM1, SM2}. First, SM eliminates the inter-channel interference (ICI) due to single antenna activation. Second, the radio frequency (RF) chain can be reduced, and thus, tight antenna synchronization can be avoided. A simplification of SM called Space Shift Keying (SSK) was proposed in \cite{5165332, 6093916}. Unlike SM, SSK obtains its symbols only from the combination of single antenna activation. However, the rate performance of these two systems is insignificant. A generalization of SM and SSK was introduced in \cite{GSM1} and \cite{5963455}, respectively. The generalization is obtained by allowing more than single antenna activation at each time. This is proposed to improve the system's achievable rate. However, besides losing the two main benefits of the conventional SM and GSK, these generalizations cannot significantly improve the achievable rate performance. In \cite{7812789}, the optimal power allocation of SM was presented. An optimization problem was formulated by deriving the mutual information of the transmit and receive signal under Gaussian distribution. However, it is unclear if we can do so due to the presence of the modulated index, which clearly does not follow Gaussian distribution.
\begin{figure}[!t]
\centering
\includegraphics [width=0.39\textwidth]{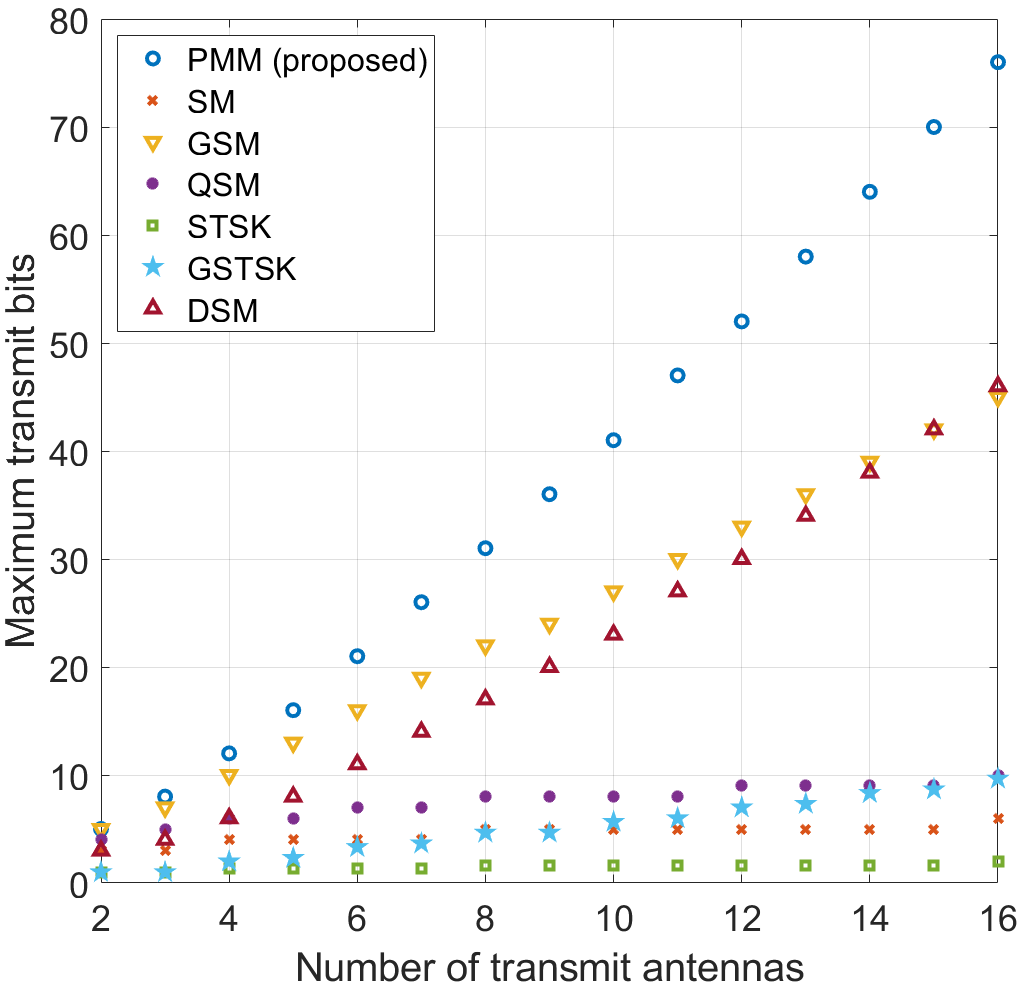} 
\caption {\rev{Comparison between our proposed system with the existing index modulation systems in the number of maximum transmit bits}}
\label{extra_bits}
\end{figure}

\rev{In \cite{5684112} and \cite{5688440}, space time shift keying (STSK) which is a matrix modulation based was proposed. The idea is to employ certain number of dispersion matrices to be modulated to represent a block of bits. The matrices are modulated along with the constellation points such that STSK has two sources of information. Although our work seems similar to STSK system, there are major differences in fact. The transmit signal of STSK is a multidimensional signal which requires multiple time instants to complete the transmission of a modulated dispersion matrix. It leads to a rate degradation since the obtained rate must be divided with the number of time instant that is used to complete transmitting one dispersion matrix. On the other hand, PMM's transmit signal is a one dimensional signal which just requires a single time instant to accommodate the transmission of a permutation matrix. In other words, PMM uses the permutation matrix to rearrange the position of power allocation coefficients from its default positions at the transmitter while STSK must receive the dispersion matrix as a full matrix. This main difference leads to distinctive performance measures. Another similar system to our work is differential spatial modulation system (DSM) in \cite{DSM} which employs permutation matrices to represent a block of bits. However, DSM is a special case of the STSK system where in DSM system permutation matrices are employed, instead of dispersion matrices. The major drawback of DSM besides it requires multiple time instants (DSM also has multidimensional transmit signal) for the signal transmission is that DSM only allows one degree of freedom since only one antenna is activated at each transmission. It is different from our work where we can utilize full degree of freedom of the MIMO system. These differences lead to significant performance differences as we show in the next sections. Moreover, the easiest way to understand the fundamental distinction of PMM with STSK and DSM is to observe the number of transmit bits in Fig. \ref{extra_bits} and Table \ref{max_txbits}.}

The focus in the early works such as \cite{DSM, 6428727, 6823072, 6328213, 8345191} was to have more \textit{"modulate-able"} symbols through antenna activation. Although some benefits are offered, as mentioned above, we argue that this method is ineffective because the potential rate from the antennas which are being inactivated is ignored. Another problem arises when we discuss the complexity of the detection scheme of these systems. Most of the authors consider maximum likelihood (ML) detection to retrieve information bits from the receive signal, such as in \cite{QSM1, 6428727}. The complexity of ML always grows exponentially with the number of inputs. For example, in this case, the complexity grows exponentially with the number of transmit antennas.

PCM system, which utilizes all-antenna activation, was originally proposed in \cite{PCM_faddli}. The basic system model and capacity analysis of PCM were discussed. A detection scheme was also proposed, and the bit error rate was simulated to evaluate the performance. The capacity was derived under the assumption of Gaussian as the input distribution. However, there is no discussion if the transmit signal can be modeled as Gaussian distribution, and furthermore, the transmit power is unconstrained at the transmitter. The system can also only work if both CSIT and CSIR are available, implying that PCM may not be competitive in practice.

\begin{table}[]
\centering
\caption{\rev{Maximum transmit bits of different index modulation systems}}
\label{max_txbits}
\begin{tabular}{|c|c|}
\hline
\rule{0pt}{2.5ex}System & Max. transmit bits \\ \hline
\rule{0pt}{2.5ex} PMM    & $A_{\mathrm{PMM}} = N\log_2(Q) + \lfloor\log_2(N!)\rfloor$ \\ \hline
\rule{0pt}{2.5ex}SM \cite{SM1}     & $A_{\mathrm{SM}} = \log_2(Q) + \lfloor\log_2(N)\rfloor$  \\ \hline
\rule{0pt}{2.5ex}GSM \cite{GSM1}    & $A_{\mathrm{GSM}} = N_a\log_2(Q) + \lfloor\log_2 {N\choose N_a} \rfloor$ \\ \hline
\rule{0pt}{2.5ex}QSM \cite{QSM1}    & $A_{\mathrm{QSM}} = \log_2(Q) + \lfloor\log_2(N^2)\rfloor$ \\ \hline
\rule{0pt}{2.5ex}STSK \cite{5684112}  & $A_{\mathrm{STSK}} = (\log_2(Q) + \log_2(L))/T$ \\ \hline
\rule{0pt}{2.5ex}DSM \cite{DSM}    & $A_{\mathrm{DSM}} = \log_2(Q) + (\lfloor\log_2(N!)\rfloor)/N$ \\ \hline
\end{tabular}
\end{table}

\subsection{Contribution}
The main idea of this paper is to propose a new structure of PCM different from \cite{PCM_faddli}. We call this new structure Permutation Matrix Modulation (PMM) instead of PCM. We provide a rigorous analysis that shows the benefits of our proposed system and how it can be more competitive compared to the others. The fundamental idea of PMM and PCM is the same, where we modulate a set of permutation matrices in addition to the conventional constellation symbols. PCM works by multiplying a permutation matrix to the singular values matrix obtained from decomposing the MIMO channel such that we can detect the altered positions of the singular values at the receiver. PMM uses a deterministic power allocation matrix known by both the transmitter and the receiver. The product of the power allocation matrix with a permutation matrix is now used, and the altered positions of the power allocations at the receiver are detected accordingly. Both systems treat a set of permutation matrices as additional symbols. PMM is also more general than PCM since PMM does not depend on the availability of CSIT because we do not need to decompose the MIMO channel. This is the reason why PMM is a more suitable name. Unlike PCM that assumes complex Gaussian as the distribution of the transmit signal, we derive an achievable rate under a Gaussian Mixture Model (GMM) distribution. We argue that GMM is an appropriate distribution for transmit signal due to the existence of the permutation matrices. We prove this mathematically in the next section.

The achievable rate derivation of SM and GSM systems under GMM distribution is available in \cite{SM_achrate}. It was obtained by finding the upper bound of the differential entropy of a GMM distribution. This upper bound is then tightened using an upper bound refinement algorithm. In this paper, we exactly follow the same methodology as \cite{SM_achrate} for deriving the achievable rate. However, we show how the upper bound refinement algorithm works and mathematically analyze why the algorithm is valid for our particular system. This discussion is missing in \cite{SM_achrate}. In tightening the upper bound, we show that our case is a special case of Salmond's clustering where we can choose any order to merge between two Gaussian mixture components from the GMM distribution. The original analysis of the upper bound refinement algorithm can be found in \cite{entr_approx} and the distance measure of two Gaussian mixture components can be found in \cite{join_Gaussmix, salmond_mixt}. \revv{Besides discussing the achievable rate obtained by having GMM distribution for the transmit signal, we also evaluate the rate of PMM through finite cardinality input (FCI) and compare the performance result with the existing systems. By using FCI method, we can analyze at what SNR our proposed system achieves its maximum transmit bits. This approach is not found in the aforementioned literature.} Furthermore, we also present an optimization problem to find the optimal power allocation to attain the optimal achievable rate of PMM. We show that the optimization problem can be solved using the well-known interior point-method.

For the performance evaluation, we compare the achievable rate of PMM with the existing systems such as SM and GSM. The result indicates that PMM outperforms SM and GSM with the same parameters. We also present the comparison between the achievable rate of PMM with optimized power allocations and generic power allocations. We show that the optimized power allocations can improve the achievable rate of PMM. Furthermore, we propose a detection scheme for PMM based on zero-forcing (ZF). We then simulate and compare the symbol error rate (SER) between ZF detection and the maximum likelihood (ML) detection. We also analyze the complexity of both detection schemes to observe the benefits and drawbacks. In short, ZF performs worse in the SER simulation but requires much less complexity compared to ML. This analysis is important from a practical point of view.

To sum up, we highlight our major contributions in this paper as follows:
\begin{itemize}
\item \rev{We continue the early work of \cite{PCM_faddli} by providing the system achievable rate performance under the GMM distribution. We also show that our system in this paper can work in the absence of CSIT, unlike in \cite{PCM_faddli}.
\item \revvv{We analyze the rate of PMM using FCI method and compare the results with the existing systems by setting the same number of maximum transmit bits.}
\item An optimization problem attaining the optimal achievable rate of PMM is discussed and evaluated. We show that despite the absence of CSIT, PMM has a close achievable rate performance to the optimal scenario. 
\item We demonstrate that the achievable rate of our proposed system outperforms the achievable rate of SM and GSM, and has a very close performance to MIMO V-BLAST. From this performance analysis, we show that full antenna activation can achieve better performance compared to partial antenna activation systems such as SM and GSM.
\item We provide the SER comparison between PMM, GSM, MIMO and SM using ML detection scheme to transmit $16$ bits per transmission. In order to achieve the same maximum transmit bits, our proposed system requires the least constellation size or number of transmit antennas. We believe that it is of the practical interest when we can achieve high performance using the least hardware requirements.
\item A new ZF-based detection for PMM is proposed, and a trade-off analysis with ML detection is given by providing the SER and complexity of each detection scheme. We show that in terms of the SER performance, ML is better than ZF. However, the complexity analysis suggests that ZF requires far less computation compared to ML.}
\end{itemize}

\subsection{Paper Outline}
The rest of this paper is organized as follows. Section II defines the system model of PMM. In Section III, we provide an analysis of the achievable rate of PMM under GMM distribution. An optimization to attain the optimal achievable rate of PMM is presented in Section IV. Section V provides detection schemes for PMM. Simulation results on the achievable rate and SER are presented in Section VI. In Section VII, the complexity analysis of the detection schemes is discussed. Finally, the major conclusions and implications are drawn in Section VIII.

\textbf{\textit{Notation:}} In the following, uppercase bold letters $\mathbf{A}$ denote matrices and lowercase bold letters $\mathbf{a}$ denote column vectors. The superscripts $(.)^{\mathrm{T}}$ and $(.)^{\mathrm{H}}$ denote transpose and conjugate-transpose, respectively. We use $\mathrm{tr}(\mathbf{A})$ and $|\mathbf{A}|$ for sum diagonal and the determinant of matrix $\mathbf{A}$, respectively, $\norm{.}$ for Frobenius norm and $\mathbb{E}\{.\}$ for the expected value.

\rev{\textbf{\textit{Reproducible research:}} Our simulation results can be reproduced using the Matlab code and data files available at: https://github.com/faddlis/Permutation-Matrix-Modulation.git}

\begin{figure*}[!t]
\centering
\includegraphics [width=0.8\textwidth]{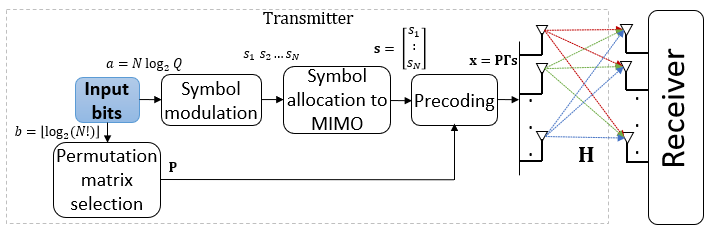} 
\caption {General system model of PMM}
\label{system_model}
\end{figure*}

\section{System Model}
We investigate a new structure in index modulation schemes called PCM initially presented in \cite{PCM_faddli}. Our proposed system is shown in Fig. \ref{system_model}. Consider a point-to-point MIMO system equipped with $N$ transmit antennas and $M$ receive antennas. The incoming information bits are divided into two blocks where the first block of length $a = N\log_2(Q)$ bits is modulated into $Q$-ary constellation symbols and the second block of length $b = \lfloor\log_2(N!)\rfloor$ bits is modulated into a permutation matrix; therefore we have a total of $a+b$ bits as the maximum transmit bits of PMM. The modulated symbols can be represented in a vector form as $\mathbf{s} = (s_1,\hdots,s_N)^{\mathrm{T}}$ where $s_i \sim \mathcal{CN}(0,1) \text{ } \forall{i}$. Element in the vector $\mathbf{s}$ indicates the position of the modulated symbol corresponding to its transmit antenna. On the other hand, the second block is used to select a permutation matrix from set $\mathcal{P} = \{\mathbf{P}_1,\hdots,\mathbf{P}_r\}$ where $r = 2^b$ is the total number of possible permutation matrices. We can see that number of the possible permutation matrix is bound to the number of transmit antennas. An example of mapping information bits to permutation matrices for $N=3$ is given in TABLE \ref{table1}. Notice that there are actually in total six unique permutation matrices for $N=3$. However, we could only use four of them for binary transmission. This is also why the floor function is used to define the number of bits in the second block $b$. We form the transmit precoded signal $\mathbf{x} \in \mathbb{C}^{N\times1}$ to convey the information bits as
\begin{equation} \label{tx_signal}
\mathbf{x} = \mathbf{P\Gamma s}
\end{equation}
where $\mathbf{P} \in \mathcal{P}$ is the modulated permutation matrix, $\mathbf{\Gamma} = \mathrm{diag}(\sqrt{\gamma_1}, \hdots, \sqrt{\gamma_N})$ is a diagonal matrix consists of the power allocated to the corresponding transmit antenna that satisfies $0 \leq \gamma_i \leq \rho \text{ } \forall{i}$, $\gamma_1 \neq \hdots \neq \gamma_N$ and $\sum_{i=1}^N \gamma_i = \rho$ where $\rho$ is the total transmit power.

\textbf{Corollary 1:} \textit{The precoded signal in (\ref{tx_signal}) satisfies the following}
\begin{equation} \label{precoded_signal}
\mathrm{tr} \left( \mathbb{E} \{ \mathbf{xx}^{\mathrm{H}}\} \right) = \mathrm{tr} \left( \mathbb{E} \{ \mathbf{P \Gamma ss}^{\mathrm{H}}\mathbf{\Gamma}^{\mathrm{H}} \mathbf{P}^{\mathrm{H}} \} \right) = \rho.
\end{equation}

\textit{Proof :} We know that $\mathbb{E}\{ \mathbf{ss}^{\mathrm{H}} \} = \mathbf{I}_N$ where $\mathbf{I}_N$ is identity matrix of length $N$, $\mathbf{\Gamma}^{\mathrm{H}} = \mathbf{\Gamma}$ since $\mathbf{\Gamma}$ is diagonal matrix and $\mathbf{P}_i^{\mathrm{H}} = \mathbf{P}_i^{-1} \text{ } \forall{i}$ since $\mathbf{P}_i$'s are unitary matrices\footnote{all permutation matrices are unitary matrix.}. Thus, matrix $\mathbf{P}$ from left-hand and right-hand sides permutes the squared elements of matrix $\mathbf{\Gamma}$ and summation of all squared elements of matrix $\mathbf{\Gamma}$ is equal to $\rho$.

Corollary 1 will later be useful in the analysis of the following sections. Notice that the precoded signal (\ref{tx_signal}) is the main difference of our proposed system in this paper with the PCM system proposed in \cite{PCM_faddli} where instead of utilizing the singular value matrix obtained from singular value decomposition (SVD) operation of the MIMO channel, we use matrix $\mathbf{\Gamma}$. Therefore, we can assume that our system can work in the absence of CSIT, unlike in PCM. Vector $\mathbf{x}$ is then sent through the MIMO channels $\mathbf{H} \in \mathbb{C}^{M\times N}$ and received as $\mathbf{y} \in \mathbb{C}^{M\times1}$ at the receiver as
\begin{equation} \label{rx_signal}
\mathbf{y} = \mathbf{Hx + n}
\end{equation}
where $\mathbf{n} \sim \mathcal{CN}(\mathbf{0},\mathbf{I}_M)$ is the resulting noise vector. Let $h_{jk}$ represents the flat-fading channel coefficient inside the channel matrix $\mathbf{H}$ drawn from Rayleigh distribution where $j$ and $k$ are the row and column indices, respectively. The entries $\mathbf{n}$ and $\mathbf{H}$ are independent and identically distributed (i.i.d) with zero mean and unit variance.

\textbf{Theorem 1:} \textit{The capacity of PMM under complex Gaussian distribution is
\begin{equation} \label{capacity_result}
C_{\mathrm{PMM}} = \log_2\left( \left| \mathbf{I}_M + \mathbf{H\tilde{\Gamma}H} \right| \right)
\end{equation}
where $\tilde{\mathbf{\Gamma}} = \mathbf{\Gamma} \mathbf{\Gamma}^{\mathrm{H}}$.}

\textit{Proof :} See APPENDIX A.

The capacity of PMM shown in (\ref{capacity_result}) is derived under the assumption that we can model the input vector $\mathbf{x}$ as Gaussian variate (e.g., $\mathbf{P}$ and $\mathbf{x}$ are Gaussian). However, it is not clear if we can do so\footnote{if matrix $\mathbf{\Gamma}$ cannot be made Gaussian distributed, $\mathbf{x}$ is not Gaussian variate.}, especially when we have finite $N$. Therefore, in order to obtain a meaningful performance analysis of our system, we will present the achievable rate by taking the assumption that vector input $\mathbf{x}$ is not Gaussian variate in the following section.

\begin{table}[!t]
\centering
\caption{Bits to permutation matrix mapping for $N=3$}
\label{table1}
\begin{tabular}{|c|c|c|}
\hline
input bits & i & $\mathbf{P}_i$ \\ \hline
00         & 1 & $\mathbf{P}_1=
			\begin{bmatrix}
			1 & 0 & 0\\ 
			0 & 1 & 0\\ 
			0 & 0 & 1
			\end{bmatrix}$     \\ \hline
01         & 2 & $\mathbf{P}_2=
			\begin{bmatrix}
			1 & 0 & 0\\ 
			0 & 0 & 1\\ 
			0 & 1 & 0
			\end{bmatrix}$    \\ \hline
10         & 3 & $\mathbf{P}_3=
			\begin{bmatrix}
			0 & 1 & 0\\ 
			1 & 0 & 0\\ 
			0 & 0 & 1
			\end{bmatrix}$    \\ \hline
11         & 4 & $\mathbf{P}_4=
			\begin{bmatrix}
			0 & 1 & 0\\ 
			0 & 0 & 1\\ 
			1 & 0 & 0
			\end{bmatrix}$    \\ \hline
\end{tabular}
\end{table}

\section{Performance Analysis of PMM}
In this section, we present the achievable rate of our proposed system. We investigate the achievable rate by first showing that the precoded vector $\mathbf{x}$ and the receive vector $\mathbf{y}$ follow GMM distribution. We then evaluate the mutual information of $\mathbf{x}$ and $\mathbf{y}$, and hence, obtain the achievable rate of PMM.

\subsection{Gaussian Mixture Model (GMM) Distribution}
We first show that every permutation matrix in the set $\mathcal{P}$ results in a unique covariance matrix. In the case of finite $N$, covariance matrix of the transmit vector $\mathbf{x}$ given that $\mathbf{P} = \mathbf{P}_i$ is
\begin{equation} \label{cov_tx}
\mathbf{C}_i = \mathbb{E} \left\{ \mathbf{xx}^{\mathrm{H}} \mid \mathbf{P} = \mathbf{P}_i \right\} = \mathbb{E} \left\{ \mathbf{P}_i \mathbf{\Gamma} \mathbf{ss}^{\mathrm{H}} \mathbf{\Gamma} \mathbf{P}_i^{-1} \right\} = \mathbb{E} \left\{\mathbf{P}_i \mathbf{\tilde{\Gamma}} \mathbf{P}_i^{-1} \right\}
\end{equation}
for $i = 1,\hdots,r$. Let $\mathcal{C} = \{\mathbf{C}_1,...,\mathbf{C}_r\}$ be the set of covariance matrices of each permutation matrix from set $\mathcal{P}$. Therefore, we know that the permutation matrix set has a one-to-one correspondence with the set of covariance matrices $\mathcal{C}$. Note that the covariance matrices are diagonal matrices of length $N$ with the same diagonal elements but in different positions. For example, the covariance matrices for $N=3$ and the permutation matrices given in TABLE \ref{table1} are shown by
\begin{equation*}
\begin{aligned}
&\mathbf{C}_1 = \begin{bmatrix}
\gamma_1 & 0 & 0\\
0 & \gamma_2 & 0\\
0 & 0 & \gamma_3
\end{bmatrix}, \text{ }
\mathbf{C}_2 = \begin{bmatrix}
\gamma_1 & 0 & 0\\
0 & \gamma_3 & 0\\
0 & 0 & \gamma_2
\end{bmatrix}, \text{ } \\
&\mathbf{C}_3 = \begin{bmatrix}
\gamma_2 & 0 & 0\\
0 & \gamma_1 & 0\\
0 & 0 & \gamma_3
\end{bmatrix}, \text{ }
\mathbf{C}_4 = \begin{bmatrix}
\gamma_2 & 0 & 0\\
0 & \gamma_3 & 0\\
0 & 0 & \gamma_1
\end{bmatrix}.
\end{aligned}
\end{equation*}

We can see that the covariance matrices are all unique and thus, distinguishable\footnote{distinguishable means that when we assume our system being free of interference and noise, we can decode the codewords (in this case the permutation matrix) successfully with probability 1.} if and only if $\gamma_1 \neq \hdots \neq \gamma_N$.

Let $f_{\mathbf{x}}^{(i)}$ be the conditional probability distribution function (pdf) of the precoded signal $\mathbf{x}$ given that $\mathbf{P} = \mathbf{P}_i$. Since the modulated symbol vector $\mathbf{s}$ is complex Gaussian, $\mathbf{x}$ is also complex Gaussian given a certain permutation matrix with zero mean and variance $\mathbf{C}_i$. We can formalize the expression as
\begin{equation} \label{pdf_x}
\begin{aligned}
f_{\mathbf{x}}^{(i)} &= p(\mathbf{x} \mid \mathbf{P} = \mathbf{P_i}) \\
&= \frac{1}{\pi^N \left| \mathbf{C}_i \right|} \exp \left( -\mathbf{x}^{\mathrm{H}} \mathbf{C}_i^{-1} \mathbf{x} \right), \text{ } i = 1,\hdots,r.
\end{aligned}
\end{equation}
On the other hand, we know that selection of the permutation matrix $\mathbf{P}_i$ is based on the incoming bitstream; therefore, each permutation matrix in the set $\mathcal{P}$ has a certain probability. Mathematically, we can write the probability mass function (pmf) of the random matrix $\mathbf{P}$ as
\begin{equation} \label{pmf_P}
p(\mathbf{P} = \mathbf{P}_i) = \alpha_i, \text{ } \forall{i}
\end{equation}
where $0 \leq \alpha_i \leq 1$ and $\sum_{i=1}^r \alpha_i = 1$.

\textbf{Theorem 2:} \textit{If the modulated symbols in vector $\mathbf{s}$ are complex Gaussian with zero mean and unit variance, \revv{and each permutation matrix has pmf as shown by (\ref{pmf_P}),} the precoded signal vector $\mathbf{x}$ follows a complex GMM distribution with pdf}
\begin{equation} \label{pdf_gmm_x}
f_{\mathbf{x}} = \sum_{i=1}^r \alpha_i \text{ } f_{\mathbf{x}}^{(i)}
\end{equation}
\textit{Proof :} See Appendix B.

We can now evaluate the covariance of random vector $\mathbf{x}$ with pdf $f_{\mathbf{x}}$. \revv{The covariance of $\mathbf{x}$} will be the same as its autocorrelation matrix since the precoded vector $\mathbf{x}$ has zero mean. The covariance of $f_{\mathbf{x}}$ is the expectation conditioned on the permutation matrix $\mathbf{P}$, and can be written as
\begin{equation} \label{cov_gx}
\mathbf{C} = \mathbb{E} \left\{ \mathbf{xx}^{\mathrm{H}} \right\} = \mathbb{E}_{\mathbf{P}}\left\{ \mathbb{E} \left\{ \mathbf{xx}^{\mathrm{H}} \mid \mathbf{P}\right\} \right\}
\end{equation}
where the conditional expectation is given by $\mathbb{E} \left\{ \mathbf{xx}^{\mathrm{H}} \mid \mathbf{P} = \mathbf{P}_i \right\} = \mathbf{C}_i$, and thus, we have
\begin{equation} \label{cov_gx2}
\mathbf{C} = \sum_{i = 1}^r \alpha_i \mathbf{C}_i.
\end{equation}

From Theorem 2, we know that the precoded signal $\mathbf{x}$ follows the GMM distribution. It is straightforward to show that the output vector $\mathbf{y}$ also follows the GMM distribution. This is due to the fact that receive vector $\mathbf{y}$ given a particular permutation matrix $\mathbf{P}_i$ has the form as follows
\begin{equation} \label{y_Pi}
\mathbf{y}_i = \mathbf{H P}_i \mathbf{\Gamma s} + \mathbf{n}.
\end{equation}
Therefore, the conditional distribution of $\mathbf{y}$ follows a complex Gaussian distribution with zero mean and variance $\mathbf{D}_i$ can be shown as
\begin{equation} \label{pdf_yi}
f_{\mathbf{y}}^{(i)} = p \left( \mathbf{y} \mid \mathbf{P} = \mathbf{P}_i \right) = \frac{1}{\pi^M \left| \mathbf{D}_i \right|} \exp \left( -\mathbf{y}^{\mathrm{H}} \mathbf{D}_i^{-1} \mathbf{y} \right),
\end{equation}
where
\begin{equation} \label{Lambda}
\mathbf{D}_i = \mathbf{HC}_i\mathbf{H}^{\mathrm{H}} + \mathbf{I}_M.
\end{equation}
Let $\mathcal{D} = \{\mathbf{D}_1,\hdots,\mathbf{D}_r\}$ be the set of covariance matrices of $\mathbf{y}$, using the same methodology as in Theorem 2, we can find the pdf of $\mathbf{y}$ to be
\begin{equation} \label{pdf_gmm_y}
f_{\mathbf{y}} = \sum_{i=1}^r \alpha_i \text{ } g_i(\mathbf{y})
\end{equation}
and the covariance of $f_{\mathbf{y}}$ is
\begin{equation} \label{cov_gy}
\mathbf{D} = \sum_{i = 1}^r \alpha_i \mathbf{D}_i.
\end{equation}

\subsection{Achievable Rate of PMM}
The mutual information of the precoded signal $\mathbf{x}$ and the receive vector $\mathbf{y}$ can be evaluated to derive the achievable rate of PMM. The mutual information is shown as follows
\begin{equation} \label{mut_inf}
I(\mathbf{x};\mathbf{y} \mid \mathbf{H}) = H(\mathbf{y} \mid \mathbf{H}) - H(\mathbf{y} \mid \mathbf{x,H})
\end{equation}
The second term in (\ref{mut_inf}) is equal to differential entropy of the noise vector $\mathbf{n}$ since it is the only remaining random variable left when $\mathbf{x}$ and $\mathbf{H}$ are given. The expression can be written as
\begin{equation} \label{mut_inf2}
H(\mathbf{y} \mid \mathbf{x,H}) = H(\mathbf{n}) = \log_2(\left| \pi e \mathbf{I}_{\mathrm{M}} \right|) = M \log_2(\pi e).
\end{equation}

On the other hand, the first term in (\ref{mut_inf}) can be evaluated as follows
\begin{equation} \label{mut_inf3}
H(\mathbf{y} \mid \mathbf{H}) = \mathbb{E}\{ -\log \left( f_{\mathbf{y}} \right) \} = - \int_{\mathbb{C}^{M}} f_{\mathbf{y}} \log \left( f_{\mathbf{y}} \right) \text{ } \mathrm{d}\mathbf{y}
\end{equation}
We can then see that the mutual information in (\ref{mut_inf}) depends on GMM distribution. Furthermore, the rate of PMM can be written as follows
\begin{equation} \label{ach_rate}
R = I(\mathbf{x};\mathbf{y} \mid \mathbf{H}) = H(\mathbf{y} \mid \mathbf{H}) - M \log_2(\pi e).
\end{equation}
Unfortunately, there is no closed-form expression for the differential entropy of a vector with a GMM pdf. This is due to the logarithm of the sum of exponential that cannot be simplified. However, we can approximate the sum logarithm in (\ref{mut_inf3}) using the multivariate Taylor-series expansion as presented in \cite{entr_approx}. Having the pdf of random vector $\mathbf{y}$ in (\ref{pdf_gmm_y}), the differential entropy of $H(\mathbf{y} \mid \mathbf{H})$ can be approximated as follows
\begin{subequations} \label{diff_ent}
\begin{align}
H(\mathbf{y} \mid \mathbf{H}) &= - \int_{\mathbb{C}^{M}} \log \left( f_{\mathbf{y}} \right) \sum_{i=1}^r \alpha_i f_{\mathbf{y}}^{(i)} \text{ } \mathrm{d}\mathbf{y} \\
&\approx - \int_{\mathbb{C}^{M}} \sum_{k=0}^L \frac{1}{k!} \left( \mathbf{y}^{\mathrm{T}} \nabla_{\mathbf{y}} \right)^k \log f_{\mathbf{y}} \mid_{\mathbf{y} = \mathbf{0}} \times \sum_{i=1}^r \alpha_i f_{\mathbf{y}}^{(i)} \mathrm{d}\mathbf{y}
\end{align}
\end{subequations}
where $L$ is the number of considered terms from the Taylor-series expansion and $\nabla_{\mathbf{y}}$ is the gradient with respect to random variable $\mathbf{y}$. The approximation is done by setting $L$ as a finite number to obtain a finite approximation, and the remaining terms are truncated. However, it is not clear if we can obtain the deviation of the approximated entropy in (\ref{diff_ent}) from its actual value as mentioned in \cite{entr_approx}. Therefore, finding the numerical result of (\ref{diff_ent}) may not be reflecting the performance of our proposed system\footnote{good approximations are usually measured from how much the approximated values are shifted at most from its actual values; therefore, derivation of the approximation deviation is required to justify if our approximation is good enough or not.}.

Another approach to obtain a meaningful performance analysis of our proposed system is to find the upper bound of rate $R$. It can be done by finding the upper bound of a GMM random vector.

\textbf{Theorem 3:} \textit{The achievable rate of PMM is}
\begin{subequations} \label{R_up}
\begin{align}
R_{\mathrm{PMM}} &= \sum_{i = 1}^r \alpha_i \left( -\log \alpha_i + \log (\pi e)^M \left| \mathbf{D}_i \right| \right) - M \log (\pi e) \\
&= \sum_{i=1}^r \alpha_i \left( \log  \frac{1}{\alpha_i}  + \log \left| \mathbf{I}_{\mathrm{M}} + \mathbf{HC}_i\mathbf{H}^{\mathrm{H}} \right| \right)
\end{align}
\end{subequations}

\textit{Proof :} Since the receive vector $\mathbf{y}$ follows a complex GMM distribution, the achievable rate of PMM follows the upper bound of differential entropy of a complex GMM distribution as shown in \cite{SM_achrate}.

The expression of the achievable rate shown in Theorem 3 tells us that there exist decoding scenarios in which we can obtain the rate $R_{\mathrm{PMM}}$ at most. Thus, the numerical results of (\ref{R_up}) can be used to analyze the performance of our proposed system.


\subsection{Upper Bound Refinement}
Note that the upper bound expression (\ref{R_up}) is not tight. However, we can tighten the upper bounds by employing the upper bound refinement algorithm as discussed in \cite{entr_approx}. The idea of this algorithm is to merge the mixture components in pairs until a single Gaussian distribution represents all the mixture components. A family of upper bounds\footnote{family of upper bound here is all possible combinations of merged and unmerged Gaussian mixture components.} will be resulted in, and the lowest value given in the family of upper bound is tighter upper bound. The complete proof that this algorithm produces tighter upper bounds is given in \cite{entr_approx}. Adapting to our case, the upper bound refinement algorithm is provided in Algorithm 1.

The problem of merging between two Gaussian mixture components shown in line 3 Algorithm 1 is one that we need to clarify. Suppose we have a mixture of $r$ Gaussian components whose pdf is given by (\ref{pdf_gmm_y}). We can merge between its two components (e.g., $\mathrm{Merge} (f_{\mathbf{y}}^{(i)},f_{\mathbf{y}}^{(j)})$) becomes a single Gaussian, as presented in \cite{join_Gaussmix}, with merged weight, mean and covariance\footnote{merged weight, mean and covariance are also often called as "moment-preserving merge".}, are respectively given by
\begin{equation} \label{merged_alpha}
\tilde{\alpha}_{ij} = \alpha_i + \alpha_j
\end{equation}
\begin{equation} \label{merged_mean}
\tilde{\pmb{\mu}}_{ij} = \alpha_{i|ij} \pmb{\mu}_i + \alpha_{j|ij} \pmb{\mu}_j
\end{equation}
\begin{equation} \label{merged_cov}
\tilde{\mathbf{C}}_{ij} = \alpha_{i|ij} \mathbf{C}_i + \alpha_{j|ij} \mathbf{C}_j + \alpha_{i|ij} \alpha_{j|ij}(\pmb{\mu}_i - \pmb{\mu}_j)(\pmb{\mu}_i - \pmb{\mu}_j)^{\mathrm{T}}.
\end{equation}
where $\alpha_{i|ij} = \alpha_i/(\alpha_i + \alpha_j)$, $\alpha_{j|ij} = \alpha_j/(\alpha_i + \alpha_j)$ and $\pmb{\mu}_i$ and $\pmb{\mu}_j$ are the means of mixture components $f_{\mathbf{y}}^{(i)}$ and $f_{\mathbf{y}}^{(j)}$, respectively. In our case, we have $\tilde{\pmb{\mu}}_{ij} = \mathbf{0} \text{ } \forall{i,j}$ and $\tilde{\mathbf{C}}_{ij} = \alpha_{i|ij} \mathbf{C}_i + \alpha_{j|ij} \mathbf{C}_j$ since we have zero mean for all our mixture components. Therefore, merging two mixture components only gives us new weight and covariance which can be computed using (\ref{merged_alpha}) and (\ref{merged_cov}), respectively. Now, we can think in what order we should merge the mixture components. One option is to use Salmond's clustering discussed in \cite{salmond_mixt} and \cite{join_Gaussmix} where we can measure the distance between $f_{\mathbf{y}}^{(i)}$ and $f_{\mathbf{y}}^{(j)}$ as follows
\begin{equation} \label{distance}
D_{\mathrm{s}}^2(i,j) = \mathrm{tr}(\tilde{\mathbf{C}}^{-1}\Delta \mathbf{W}_{ij})
\end{equation}
where
\begin{equation} \label{delta_W}
\Delta \mathbf{W}_{ij} = \frac{\alpha_i \alpha_j}{\alpha_i + \alpha_j}(\pmb{\mu}_i - \pmb{\mu}_j)(\pmb{\mu}_i - \pmb{\mu}_j)^{\mathrm{T}},
\end{equation}
and $\tilde{\mathbf{C}}$ is the overall covariance. Having the distance (\ref{distance}), we must compare all the distance of the mixture components and choose two mixture components with the smallest distance to be merged first.

\textbf{Corollary 2:} \textit{We can choose arbitrary order to merge the Gaussian mixture components of $f_{\mathbf{y}}$ to obtain tighter upper bounds (\ref{R_up}).}

\textit{Proof :} Since we have zero mean for all the Gaussian mixture components, we have $\Delta \mathbf{W}_{ij} = \mathbf{0} \text{ } \forall{i,j}$, therefore, we always have $D_{\mathrm{s}}^2(i,j) = 0 \text{ } \forall{i,j}$. In other words, merging any of two components of $f_{\mathbf{y}}$ always results in zero distance which means any merging order will always result in the same tighter upper bound.

We know from Corollary 2 that our "Merge" is a special case of Salmond's clustering. Thus, we can merge the mixture components, for example, in the order shown by Algorithm 1.

\begin{algorithm}[!t]
    \SetAlgoLined
    \DontPrintSemicolon
    \KwInput{$\{r\}$, $\{\alpha_i, \text{ } i=1,\hdots,r\}$ and $\mathcal{C}$}
    \KwInitialization{Set $\tilde{f}_{\mathbf{y}} \leftarrow f_{\mathbf{y}}$,$\text{ } \tilde{\alpha} \leftarrow 0$}
    Compute $R_{\mathrm{PMM}}^{(1)}$ using (\ref{R_up})\;
     \For{$k = 2,\hdots,r$}{
      $\tilde{f}_{\mathbf{y}} \leftarrow$ Merge$(f_{\mathbf{y}}^{(k-1)},f_{\mathbf{y}}^{(k)})$ \;
      $\tilde{\alpha} \leftarrow \alpha_{k-1}+\alpha_{k}$ \;
      Compute $R_{\mathrm{PMM}}^{(k)}$ using $\tilde{f}_{\mathbf{y}}$ and $\tilde{\alpha}$ \;
      $R_{\mathrm{tight}} \leftarrow \min(R_{\mathrm{PMM}}^{(k-1)},R_{\mathrm{PMM}}^{(k)})$ \;
      $f_{\mathbf{y}}^{(k)} \leftarrow \tilde{f}_{\mathbf{y}}$ \;
      $\alpha_k \leftarrow \tilde{\alpha}$
     }
     \KwOutput{$R_{\mathrm{tight}}$}
     \caption{Upper Bound Refinement Algorithm}
\end{algorithm}

\textit{\textbf{Remark 1:} Salmond's clustering is an approximation method to find the distance of Gaussian mixture components. There are several other methods, such as Kullback-Leibler discrimination and Runnall's reduction method in \cite{join_Gaussmix}. It is important to note that all these methods do not guarantee to produce the tightest upper bound. We chose Salmond's clustering because it requires cheap computation and has a reasonable trade-off with precision.}

\revvv{\subsection{Achievable Rate via Finite-Cardinality Input (FCI)}
In the previous section, we have shown the achievable rate expression of our proposed system by considering GMM distribution. It implies that the achievable rate shown in (\ref{R_up}) was evaluated using continuous distributions as the input. Another method that we can also use to observe the achievable rate of PMM is to consider finite input distribution. In \cite{dcmc}, this method is called discrete-input continuous-output memoryless channel (DCMC) and in \cite{proakis2008digital}, it is called the cut-off rate. The idea is to employ a finite set of codewords and define the probability of each codeword being transmitted. Using this method, the achievable rate will be restricted on the maximum transmit bits as shown in TABLE \ref{max_txbits}. It means that in our case, we include number of antenna and constellation size as factors in evaluating the achievable rate via FCI.

In evaluating the achievable rate via FCI, we first define the conditional probability of receiving the transmitted signal (\ref{tx_signal}) given that $\mathbf{P} \in \mathcal{P}$ and $\mathbf{s} = ({s}_1, \hdots, {s}_N)^\mathrm{T}$ where $s_i \in \mathcal{Q}$ and $\mathcal{Q}$ is the set containing all $Q$-ary constellation symbols. For simplicity, let us then define $\mathbf{x}_m$ for $m = 1,\hdots,V$ as the $m$-th transmit signal from the total of $V = Q^N$ possible transmit signal combinations. Thus, the achievable rate of PMM can be evaluated by \cite[Ch. 6.8]{proakis2008digital}
\begin{equation} \label{ARviaFCI0}
\begin{aligned}
R_{PMM}^{FCI} = &\max_{p(\mathbf{x}_1),\hdots,p(\mathbf{x}_V)} \\
&-\log_2 \left( \int_{-\infty}^{\infty} \left( \sum_{v=1}^V p(\mathbf{x}_v) \sqrt{p(\mathbf{y} | \mathbf{x}_v,\mathbf{H})} \right)^2 d\mathbf{y} \right).
\end{aligned}
\end{equation}
To solve (\ref{ARviaFCI0}), we can use the inequalities
\begin{subequations} \label{ineqFCI}
\begin{align}
\int_{-\infty}^{\infty} &\left( \sum_{v=1}^V \sqrt{p(\mathbf{y} | \mathbf{x}_v,\mathbf{H})} \right)^2 d\mathbf{y} \geq \sum_{v=1}^V \int_{-\infty}^{\infty} p(\mathbf{y} | \mathbf{x}_v,\mathbf{H}) d\mathbf{y} \\
&= \sum_{v=1}^V \mathbb{E}\left\{ \sum_{w=1}^V \exp\left( \frac{-\mathrm{SNR}}{4} \norm{\mathbf{H}(\mathbf{x}_v - \mathbf{x}_w)}_F^2 \right) \right\},
\end{align}
\end{subequations}
where we can solve the expectation in (\ref{ineqFCI}b) by finding the average power gain assuming i.i.d Rayleigh fading of each element in the channel $\mathbf{H}$ as shown in \cite[Ch. 14.4]{proakis2008digital}, so that we have
\begin{equation} \label{solve_exp}
\begin{aligned}
\exp&\left( \frac{-\mathrm{SNR}}{4} \norm{\mathbf{H}(\mathbf{x}_v - \mathbf{x}_w)}_F^2 \right) \leq \\
&\frac{1}{\left( \left| \mathbf{I}_N + \frac{\mathrm{SNR}}{4} (\mathbf{x}_v - \mathbf{x}_w)(\mathbf{x}_v - \mathbf{x}_w)^\mathrm{H} \right| \right)^M}.
\end{aligned}
\end{equation}
Using the fact that (\ref{ARviaFCI0}) is maximized when we have a uniform distribution on the input codewords (e.g., $p(\mathbf{x}_v) =\frac{1}{V} \text{ } \forall{v}$) and by substituting (\ref{ineqFCI}) and (\ref{solve_exp}) to (\ref{ARviaFCI0}), we have
\begin{equation} \label{ARviaFCI}
\begin{aligned}
&R_{PMM}^{FCI} = \log_2(V) \\
&- \log_2 \left( \sum_{v=1}^V \sum_{w=1}^V \frac{1}{\left( \left| \mathbf{I}_N + \frac{\mathrm{SNR}}{4} (\mathbf{x}_v - \mathbf{x}_w)(\mathbf{x}_v - \mathbf{x}_w)^\mathrm{H} \right| \right)^M} \right),
\end{aligned}
\end{equation}
where $\mathrm{SNR} = \frac{\rho}{N0}$ is the signal to noise power ratio.
}


\section{Optimal Achievable Rate with CSIT}
Suppose we know the channel matrix $\mathbf{H}$ at the transmitter, we can have $\mathbf{H} = \mathbf{U \Lambda V}^{\mathrm{H}}$ by decomposing $\mathbf{H}$ using the well-known singular value decomposition (SVD) where $\mathbf{U}$ and $\mathbf{V}$ are both unitary matrices, and $\mathbf{\Lambda} = \mathrm{diag}(\lambda_1,\hdots,\lambda_N)$ is a diagonal matrix containing the singular values of $\mathbf{H}$. We can reform the precoded signal as
\begin{equation} \label{x_csit}
\mathbf{x}_{\mathrm{csit}} = \mathbf{V P \Gamma s},
\end{equation}
and obtain the receive signal as
\begin{equation} \label{y_csit}
\mathbf{y}_{\mathrm{csit}} = \mathbf{Hx}_{\mathrm{csit}} + \mathbf{n} = \mathbf{U \Lambda P \Gamma s} + \mathbf{n}.
\end{equation}
Finally, we obtain a parallel channel by multiplying singular matrix $\mathbf{U}$ to $\mathbf{y}_{\mathrm{csit}}$ as shown by
\begin{equation} \label{post_proc_csit}
\tilde{\mathbf{y}}_{\mathrm{csit}} = \mathbf{U}^{\mathrm{H}} \mathbf{y}_{\mathrm{csit}} = \mathbf{\Lambda P \Gamma s} + \tilde{\mathbf{n}} = \sum_{k=1}^M \sqrt{\gamma_{k(i)}} \lambda_k s_{k(i)} + \tilde{n}_k
\end{equation}
where $\tilde{\mathbf{n}} = \mathbf{U}^{\mathrm{H}} \mathbf{n} = (\tilde{n}_1,\hdots,\tilde{n}_M)^{\mathrm{T}}$ and $i\in \{1,\hdots,N\}$. Note that the noise statistics  in (\ref{post_proc_csit}) is preserved since unitary matrix $\mathbf{U}$ only rotates the noise vector. We write the index $(i)$ to show that the position of $\gamma_k$ may be different from its original position at the transmitter. This is due to the multiplication of the permutation matrix $\mathbf{P}$ may alter the position of $\gamma_k$ when received at the receiver. So, the index $(i)$ is there to keep the original position of $\gamma_k$ tracked at the receiver. Following the same analysis given in Section III, the achievable rate can be written as
\begin{subequations} \label{R_csit}
\begin{align}
R_{\mathrm{csit}} &= \sum_{j=1}^r \alpha_j \left( \log  \frac{1}{\alpha_j}  + \log \left| \mathbf{I}_{\mathrm{M}} + \mathbf{\Lambda} \mathbf{P}_j \mathbf{\Gamma} \mathbf{P}_j \mathbf{\Lambda} \right| \right) \\
&= \sum_{j=1}^r \alpha_j \left( \log  \frac{1}{\alpha_j}  + \sum_{k=1}^M \log \left(1 + \lambda_k^2 \gamma_{k(i)} \right) \right).
\end{align}
\end{subequations}

We can attain the optimal achievable rate $R_{\mathrm{csit}}^*$ by finding the optimal power allocation $\mathbf{\Gamma}^*$ by assuming that we know the moment-preserving merge that gives the tight upper bound. It is sensible to assume that $\alpha_i$ follows uniform distribution, $\alpha_i = \frac{1}{r} \text{ } \forall{i}$, since typically we do not have much control over the incoming bitstreams. Let $\mathbf{C}_{\mathrm{tight}}$ and $\alpha_{\mathrm{tight}}$ be the covariance and weight of the merged Gaussian mixture components that give the tight upper bound, respectively. The optimization to obtain $R_{\mathrm{tight}}^*$ can be formulated as follows
\begin{equation} \label{optim_Rcsit}
\begin{aligned}
& \underset{ \gamma_j \in [0,\rho], \text{ } j=1,\hdots,N}{\text{maximize}} \text{ }\text{ } R_{\mathrm{tight}} = X + Y \\
& \text{subject to}
\text{ } \text{ } \sum_{j=1}^N \gamma_j = \rho, \\
\end{aligned}
\end{equation}
where $X$ and $Y$ are the merged and unmerged mixture components, respectively. For example, we obtain the tight upper bound of $R_{\mathrm{csit}}$ by merging the mixture components from $f_{\mathbf{y}}^{(1)}$ to $f_{\mathbf{y}}^{(u)}$ where $u\leq r$, which means we have unmerged components $f_{\mathbf{y}}^{(u+1)},\hdots,f_{\mathbf{y}}^{(r)}$. Thus, we have $\mathbf{C}_{\mathrm{tight}} = \frac{1}{u}\sum_{j = 1}^u \mathbf{P}_j \mathbf{\Gamma} \mathbf{P}_j$ and $\alpha_{\mathrm{tight}} = \frac{u}{r}$. In this case, we can write $R_{\mathrm{tight}}$ to be
\begin{equation} \label{R_tight}
\begin{aligned}
R_{\mathrm{tight}} = &\underbrace{\alpha_{\mathrm{tight}} \left( \log \frac{1}{\alpha_{\mathrm{tight}}} + \log \left| \mathbf{I}_{\mathrm{M}} + \mathbf{\Lambda} \mathbf{C}_{\mathrm{tight}} \mathbf{\Lambda} \right| \right)}_\text{$X$} + \\ &\underbrace{\sum_{j=u+1}^r \alpha_j \left( \log  \frac{1}{\alpha_j}  + \log \left| \mathbf{I}_{\mathrm{M}} + \mathbf{\Lambda} \mathbf{P}_j \mathbf{\Gamma} \mathbf{P}_j \mathbf{\Lambda} \right| \right)}_\text{$Y$}.
\end{aligned}
\end{equation}
Note that $Y = 0$ when $R_{\mathrm{tight}}$ is obtained from merging all the mixture components becomes a single component. We can see that all matrix terms inside the determinant in (\ref{R_tight}) are diagonal matrices which can be written as summation of logarithmic function similar to (\ref{R_csit}b). Since $R_{\mathrm{tight}}$ is a concave function within the feasible solution, the optimization problem in (\ref{optim_Rcsit}) can be reformed as follows
\begin{equation} \label{optim_Rcsit2}
\begin{aligned}
& \underset{ \gamma_j, \text{ } j=1,\hdots,N}{\text{minimize}} \text{ }\text{ } -R_{\mathrm{tight}}\\
& \text{subject to}
\text{ } \text{ } 0 \leq \gamma_j \leq \rho, \text{ } j = 1,\hdots,N\\
&\text{ } \text{ } \text{ } \text{ } \text{ } \text{ } \text{ } \text{ } \text{ } \text{ } \text{ } \text{ } \sum_{j=1}^N \gamma_j = \rho. \\
\end{aligned}
\end{equation}

\textbf{Corollary 3:} \textit{The optimization problem in (\ref{optim_Rcsit2}) can be efficiently solved using interior-point methods.}

\textit{Proof :} By noting that the constant $\lambda_j, \gamma_j \text{ } \forall{j}$ are non-negative, all terms inside the determinant function are non-negative. Therefore, the objective function $R_{\mathrm{tight}}$ is a strictly concave function within the feasible solution and twice differentiable. Furthermore, all constraints are linear, and the solution is always feasible. This directly follows the standard form of interior-point methods with inequality and equality constrained minimization problems in \cite[Ch. 11]{boyd2004convex}.

\section{Detection Schemes for PMM}
By setting $\gamma_1 \neq \hdots \neq \gamma_N$, we have shown that the permutation matrices from the set $\mathcal{P}$ are distinguishable, and therefore, can be detected at the receiver. The problem in this section is to design detection schemes (to detect random matrix $\mathbf{P}$ and modulated symbol $\mathbf{s}$) that can work for PMM and analyze its performance.

\subsection{Maximum Likelihood (ML) Detection}
One option is to perform ML detection, where we examine all possible combinations and find the best combination among the others. Using ML detection for PMM, we can formulate the problem as follows

\begin{equation} \label{ML}
\begin{aligned}
[\hat{\mathbf{P}},\hat{{s}}_1,\hdots,\hat{{s}}_N] &= \argmin_{\hat{\mathbf{P}} \in \mathcal{P}, \text{ } \hat{s}_k \in \mathcal{Q}, \text{ }  k = 1,\hdots,N} \norm{\mathbf{y} - \mathbf{H \hat{P} \Gamma \hat{s}}}^2 \\ 
&= \argmin_{\hat{\mathbf{P}} \in \mathcal{P}, \text{ } \hat{s}_k \in \mathcal{Q}, \text{ }  k = 1,\hdots,N} g(\mathbf{\hat{P}},\mathbf{\hat{s}})
\end{aligned}
\end{equation}
where $\hat{\mathbf{P}}$ and $\hat{\mathbf{s}} = (\hat{s}_1, \hdots, \hat{s}_N)^{\mathrm{T}}$ are the detected permutation matrix and modulated symbols at the receiver, respectively, and $\mathcal{Q}$ is the set containing all $Q$-ary constellation symbols. We can view the problem in (\ref{ML}) as finding the best combination from set $\mathcal{P}$ and $\mathcal{Q}$ that minimizes the cost function $g$.

\revv{Let $V = Q^N$ be the number of all possible combinations from input $\hat{\mathbf{P}}$ and $\hat{\mathbf{s}}$ to form the transmitted signal $\mathbf{x}_i$, we can simplify the cost function $g$ in the problem (\ref{ML}) given that $\mathbf{x}_i$ was sent to be
\begin{equation} \label{func_g}
g(\mathbf{\hat{P}},\mathbf{\hat{s}}) = \norm{\mathbf{n} + \mathbf{H} (\mathbf{x}_i - \mathbf{x}_v)}^2 = \norm{\mathbf{n} + \hat{\mathbf{e}}_v}^2, \text{ } v = 1,\hdots,V
\end{equation}
where $\hat{\mathbf{e}}_v$ is the $v$-th combination of input $\hat{\mathbf{P}}$ and $\hat{\mathbf{s}}$. Thus, when $v = i$, we have $\hat{\mathbf{e}}_v = \mathbf{0}$. We know that successful detection is attained when
\begin{equation} \label{cor_ML}
\norm{\mathbf{n}} < \norm{\mathbf{n} + \hat{\mathbf{e}}_v} \text{ } \forall{v \neq i}.
\end{equation}
We can write the probability of successful detection as follows
\begin{equation} \label{prob_ML}
P_{\mathrm{c}}^{\mathrm{ML}} = p(\norm{\mathbf{n}} < \norm{\mathbf{n} + \hat{\mathbf{e}}_v} \text{ } \forall{v \neq i}).
\end{equation}}
And finally, we have the probability of incorrect detection as shown by
\begin{equation} \label{prob_inc_ML}
P_\mathrm{err}^{\mathrm{ML}} = 1 - P_{\mathrm{c}}.
\end{equation}


\textit{\textbf{Remark 2:} In the extreme case, when $\norm{\mathbf{n}} = 0$, we can achieve $100\%$ correct detection (e.g., $P_{\mathrm{c}} = 1$). This means that the expression in (\ref{cor_ML}) tells us that we can maximize the probability of successful detection by minimizing $\norm{\mathbf{n}}$. One way to achieve this is to maximize the receive SNR. We will show this later in the simulation results section.} 

\subsection{Zero-Forcing (ZF) Detection}
The detection scheme using ML is based on exhaustive search, which leads to high complexity\footnote{complexity here means the computational complexity.}. Using ZF, we can design a detection scheme to retrieve the information conveyed by both $\mathbf{s}$ and $\mathbf{P}$. The basic idea of ZF is to remove the inter-stream interference. It is done by projecting the received signal $\mathbf{y}$ onto the subspace orthogonal to the one spanned by the vector columns of $\mathbf{H}$. Initially, the projection is performed at each stream. However, there is a well-known explicit formula where we can decorrelate all streams at once by forming the pseudoinverse of channel matrix $\mathbf{H}$ \cite{tse}, the pseudoinverse matrix can be defined as follows
\begin{equation} \label{matrix_ZF}
\mathbf{H}_{\mathrm{ZF}} = (\mathbf{H}^{\mathrm{H}} \mathbf{H})^{-1} \mathbf{H}^{\mathrm{H}}
\end{equation}
where in the special case when $N=M$, $\mathbf{H}_{\mathrm{ZF}} = \mathbf{H}^{-1}$.

We then perform a post-precessing by multiplying the ZF matrix with the received signal, as shown by
\begin{equation} \label{post_proc_ZF}
\tilde{\mathbf{y}} = \mathbf{H}_{\mathrm{ZF}} \mathbf{y} = \mathbf{x} + \mathbf{H}_{\mathrm{ZF}} \mathbf{n}.
\end{equation}
Note that applying ZF as the post-processing results in a colored noise $\mathbf{H}_{\mathrm{ZF}} \mathbf{n}$. In other words, the statistic of the noise $\mathbf{n}$ is changed due to the application of ZF. We can detect the permutation matrix using the following expression
\begin{equation} \label{detect_perm_M}
\hat{\mathbf{P}} = \argmax_{\hat{\mathbf{P}} \in \mathcal{P}} \norm{\mathbf{\Gamma\hat{P}}^{\mathrm{H}} \tilde{\mathbf{y}}}^2 = \argmax_{\hat{\mathbf{P}} \in \mathcal{P}} h(\hat{\mathbf{P}}).
\end{equation}
The cost function $h$ can be rewritten to be
\begin{equation} \label{ZF_err}
\begin{aligned}
h(\hat{\mathbf{P}}) &= \norm{\mathbf{\Gamma\hat{P}}^{\mathrm{H}} \mathbf{x} + \mathbf{\Gamma\hat{P}}^{\mathrm{H}}\mathbf{H}_{\mathrm{ZF}} \mathbf{n}}^2 \\
&= \norm{\mathbf{\Gamma\hat{P}}^{\mathrm{H}} \mathbf{x} + \pmb{\epsilon}_j}^2, \hat{\mathbf{P}} \in \mathcal{P} \text{ } \mathrm{and} \text{ } j=1,\hdots,r
\end{aligned}
\end{equation}
where $\pmb{\epsilon}_j$ is the error resulting from multiplying the $j$-th permutation matrix from set $\mathcal{P}$. Let $l$ be the correct permutation matrix index. Therefore, we have the correct detection of the permutation matrix when
\begin{equation} \label{ZF_PM_correct}
\norm{\mathbf{\Gamma{P}}_l^{\mathrm{H}} \mathbf{x} + \pmb{\epsilon}_l} > \norm{\mathbf{\Gamma{P}}_j^{\mathrm{H}} \mathbf{x} + \pmb{\epsilon}_j}, j\neq l.
\end{equation}
The probability of correct detection of the permutation matrix can be written as
\begin{equation} \label{prob_PM_correct}
P_c^{(\mathrm{PM})} = p\left( \norm{\mathbf{\Gamma{P}}_l^{\mathrm{H}} \mathbf{x} + \pmb{\epsilon}_l} > \norm{\mathbf{\Gamma{P}}_j^{\mathrm{H}} \mathbf{x} + \pmb{\epsilon}_j}, j\neq l \right).
\end{equation}

After detecting the permutation matrix, we then multiply inverse of the detected permutation matrix with $\tilde{\mathbf{y}}$ as shown by
\begin{equation} \label{detect_symb}
\tilde{\mathbf{y}}_{\hat{\mathbf{P}}} = \mathbf{\hat{P}}^{\mathrm{H}} \tilde{\mathbf{y}},\text{ } \hat{\mathbf{P}} \in \mathcal{P}.
\end{equation}
Suppose $\hat{\mathbf{P}}$ is correct, the equation (\ref{detect_symb}) is purposed to restore the constellation symbols into their original positions. Finally, we can retrieve the information bits from (\ref{detect_symb}) using 
\begin{equation} \label{detect_symb2}
\hat{s}_k = \argmin_{\hat{s} \in \mathcal{Q}} \norm{\tilde{y}_k - \hat{s}}^2
\end{equation}
where $\tilde{y}_k$ is the $k$-th element of vector $\tilde{\mathbf{y}}_{\hat{\mathbf{P}}}$. When we assume that the permutation matrix is correctly detected, we can write the probability of correct detection of the symbol at the $k$-th position as
\begin{equation} \label{prob_corr_symbol}
\begin{aligned}
&P_{\mathrm{c}}^{(s_k)} = \\
&p\left( \norm{\gamma_k s_k + \mathbf{h}_{\mathrm{ZF}}^{(k)}\mathbf{n}} < \norm{\gamma_k \hat{s} + \mathbf{h}_{\mathrm{ZF}}^{(k)}\mathbf{n}}, \hat{s} \in \mathcal{Q} \text{ } \mathrm{and} \text{ } s_k\neq \hat{s} \right).
\end{aligned}
\end{equation}
where $\mathbf{h}_{\mathrm{ZF}}^{(k)}$ is the $k$-th row of matrix $\mathbf{H}_{\mathrm{ZF}}$. We can have the probability of overall correct detection as follows
\begin{equation} \label{prob_corr_ZF}
P_{\mathrm{c}}^{\mathrm{ZF}} = P_c^{(\mathrm{PM})} \prod_{k=1}^M P_{\mathrm{c}}^{(s_k)}.
\end{equation}
And finally, the probability of incorrect detection can be written as follows
\begin{equation} \label{prob_incorr_ZF}
P_{\mathrm{e}}^{\mathrm{ZF}} = 1 - P_{\mathrm{c}}^{\mathrm{ZF}}.
\end{equation}
We must mention that there are certain limitations when we use ZF detection. The first explicit limitation is that $N \leq M$, because the inter-stream interference removal is successful if the $i$-th column vector of $\mathbf{H}$ is not a linear combination of the other column vectors of $\mathbf{H}$. In other words, if there are more streams than the dimension of the received signal (e.g., $N > M$), the ZF detection will not be successful.

\begin{algorithm}[!t]
    \SetAlgoLined
    \DontPrintSemicolon
    \KwInput{MIMO channel matrix $\mathbf{H}$ and receive signal $\mathbf{y}$}
    Generate the pseudoinverse matrix $\mathbf{H}_{\mathrm{ZF}}$ using (\ref{matrix_ZF})\;
    Multiply $\mathbf{H}_{\mathrm{ZF}}$ with the receive signal $\mathbf{y}$ to obtain $\tilde{\mathbf{y}}$\;
    Detect the permutation matrix using (\ref{detect_perm_M})\;
    Multiply the detected permutation matrix $\hat{\mathbf{P}}$ with $\tilde{\mathbf{y}}$\;
    Detect the transmit constellation symbols using (\ref{detect_symb2})\;
     \KwOutput{$\hat{\mathbf{P}}$, $s_1,\hdots,s_N$}
     \caption{ZF Detection Scheme Algorithm for PMM}
\end{algorithm}

The second limitation is that $|s_1| = \hdots  = |s_N|$. This is due to the permutation matrix detection process shown in (\ref{detect_perm_M}). It exactly follows the same reason in \cite[Theorem 1]{PCM_faddli} where when we suppose $\norm{\mathbf{n}} = 0$, the equation (\ref{detect_perm_M}) only produces the correct result either when $|\gamma_1| = \hdots  = |\gamma_N|$ or $|s_1| = \hdots  = |s_N|$. Therefore, our only option is to satisfy the second constraint since the first one cannot be satisfied in order to fulfill the distinguishable constraint of the permutation matrix. We can satisfy the second one by for example employing the phase shift keying (PSK) modulation.

ZF detection scheme for PMM is our proposed detection scheme to reduce the complexity resulting in ML detection. This makes ZF detection becomes one of our contributions in this paper. The ZF detection scheme for PMM is summarized in Algorithm 2.

\textit{\textbf{Remark 3:} The fundamental difference between the detection using ML and ZF is that ML is a joint detection scheme while ZF detects the symbols step-by-step, where the detection of constellation symbols depends on the correctness of the permutation matrix detection. This can be easily verified from (\ref{prob_ML}) and (\ref{prob_corr_ZF}). Note that the probability of error shown by (\ref{prob_inc_ML}) and (\ref{prob_incorr_ZF}) are the probability where the permutation matrix and all constellation symbols are incorrectly detected. This is not generally true since, in some cases, we could still have several symbols correctly detected even though some symbols are incorrectly detected.}


\section{Simulations}
We provide the simulation results of the achievable rate and symbol error rate (SER) to support our analysis presented in the previous sections. In general, we employ Rayleigh flat-fading channel and AWGN with zero means and unit variances for all simulations. We also assume that perfect CSIR is known for all simulations. All the achievable rates are evaluated using the upper bound refinement algorithm presented in Algorithm 1 and simulated under the GMM as the input distribution. Throughout this simulation, we use TABLE \ref{pow_alloc} for the power allocation of PMM (unless specified otherwise). This power allocation is generic by only considering that we must set different power allocations at each antenna for PMM. GSM employs equal power allocation, $\frac{1}{N_{\mathrm{act}}} \rho$ where $N_{\mathrm{act}}$ is the number of activated antennas at each transmission. Furthermore, SM employs maximum power for the activated antenna. Other related parameters will be specified at each simulation.

\begin{table}[]
\centering
\caption{Power allocation of PMM}
\label{pow_alloc}
\begin{tabular}{|c|c|c|c|c|c|c|}
\hline
$N$ & $\gamma_1$ & $\gamma_2$ & $\gamma_3$ & $\gamma_4$ & $\gamma_5$ & $\gamma_6$ \\ \hline
$3$ & $0.39\rho$ & $0.33\rho$ & $0.28\rho$ & $-$ & $-$   & $-$    \\ \hline
$4$ & $0.34\rho$ & $0.28\rho$ & $0.22\rho$ & $0.16\rho$ & $-$   & $-$    \\ \hline
\revv{$5$} & \revv{$0.32\rho$} & \revv{$0.26\rho$} & \revv{$0.20\rho$} & \revv{$0.14\rho$} & \revv{$0.08\rho$}   & \revv{$-$}    \\ \hline
$6$ & $0.27\rho$ & $0.23\rho$ & $0.19\rho$ & $0.14\rho$ & $0.1\rho$ & $0.07\rho$ \\ \hline
\end{tabular}
\end{table}

We compare our proposed system with two existing index modulation systems, SM and GSM. The analysis about the existing systems is discussed in \cite{SM_achrate}. From the achievable rate perspective, the main difference between the existing systems with our proposed system is the modulated indices\footnote{we refer modulated indices as additional symbols that are not obtained from constellation symbols.}. In short, SM modulates a single antenna as additional symbols to convey information bits. Thus, one antenna is activated at each transmission, and the combination of different antenna activation is the symbol used to send more information bits. While GSM allows any number of antennas to be activated, GSM obtains its additional symbols from combining those multiple antennas activation. As presented in \cite{SM_achrate}, SM's achievable rate under GMM distribution can be computed as follows
\begin{equation} \label{SM_rate}
R_{\mathrm{SM}} =  \sum_{i=1}^t \alpha_i^{\mathrm{SM}} \left( \log  \frac{1}{\alpha_i^{\mathrm{SM}}}  + \log \left| \mathbf{I}_{\mathrm{M}} + \mathbf{HC}_i^{\mathrm{SM}}\mathbf{H}^{\mathrm{H}} \right| \right)
\end{equation}
where $t$ and $\mathbf{C}_i^{\mathrm{SM}}$ are the total number of SM's modulated indices and the covariance of SM when employing the $i$-th modulated index, respectively. And for the achievable rate for GSM is
\begin{equation} \label{GSM_rate}
R_{\mathrm{GSM}} =  \sum_{i=1}^v \alpha_i^{\mathrm{GSM}} \left( \log  \frac{1}{\alpha_i^{\mathrm{GSM}}}  + \log \left| \mathbf{I}_{\mathrm{M}} + \mathbf{HC}_i^{\mathrm{GSM}}\mathbf{H}^{\mathrm{H}} \right| \right)
\end{equation}
where $v$ and $\mathbf{C}_i^{\mathrm{GSM}}$ are the total number of GSM's modulated indices and the covariance of GSM when employing the $i$-th modulated index.

\begin{figure}[!t]
\centering
\includegraphics [width=0.4\textwidth]{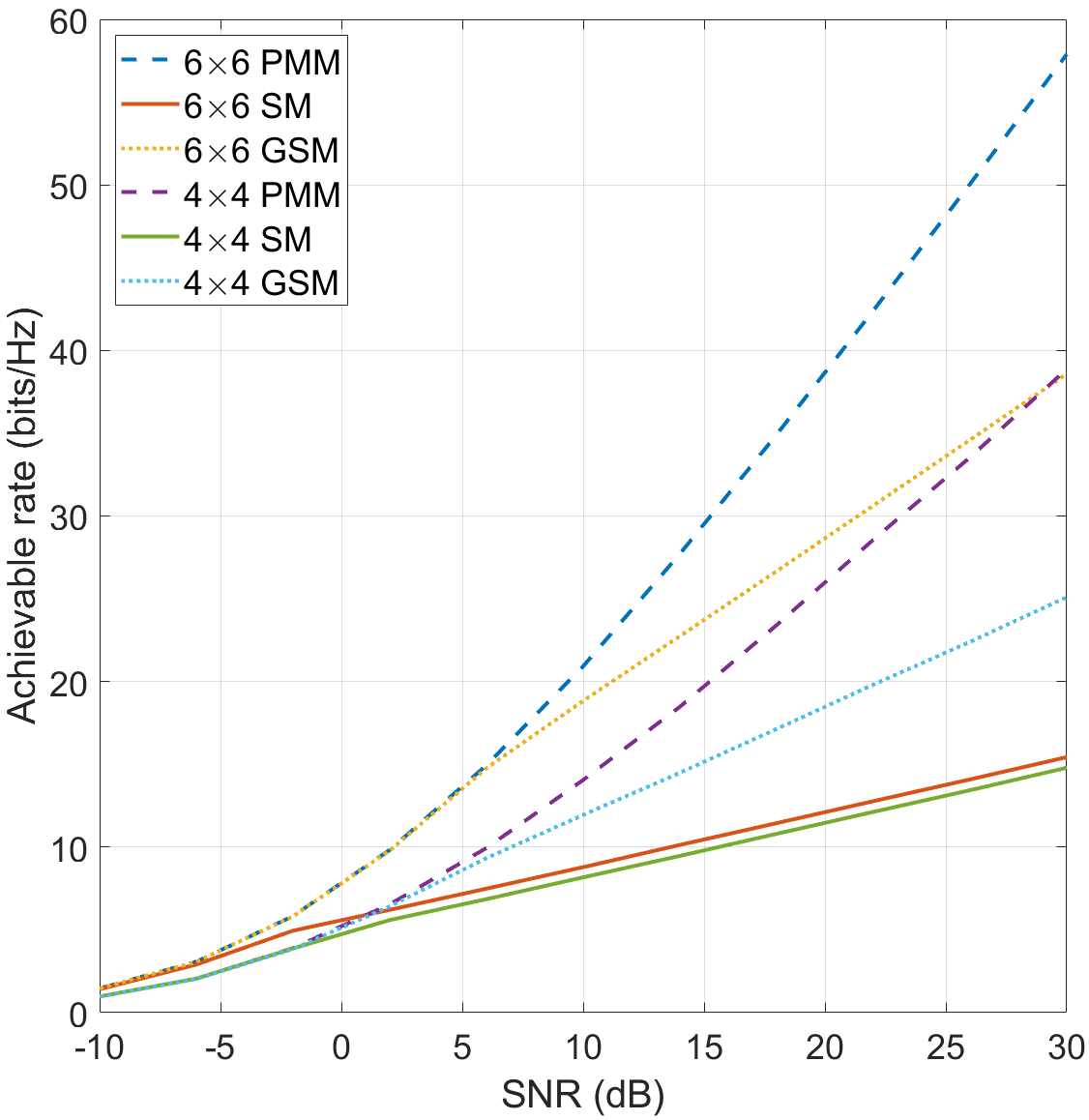} 
\caption {Achievable rate of PMM, SM and GSM with different antenna settings}
\label{rate1}
\end{figure}

See that (\ref{SM_rate}) and (\ref{GSM_rate}) are similar to the achievable rate of PMM shown in (\ref{R_up}). The only difference is the covariance matrices of each system. For SM, since it only allows one antenna activation at each time, each covariance matrix contains all zeros except one diagonal element containing maximum power, which indicates the antenna that is being activated. For GSM, the covariance matrices are also all zeros except the diagonal elements that indicate which antennas combination are being activated with the trace of each covariance matrix is equal to maximum power.

\subsection{Achievable rate}
The achievable rates of PMM, SM and GSM are presented in Fig. \ref{rate1} with $4 \times 4$ and $6 \times 6$ antenna settings. We can see that PMM outperforms both SM and GSM over the whole SNR range. Increasing the antenna setting from $4 \times 4$ to $6 \times 6$ improves in average $49.4 \%$ of the rate for PMM, $50.5 \%$ for GSM and $15.3 \%$ for SM. This reflects that PMM and GSM gain more modulated indices than SM when the antenna setting increases. In fact, SM does not obtain any additional modulated indices from increasing the antenna setting in Fig. \ref{rate1} since SM is bounded to $2^N$ in order to gain usable modulated indices for binary transmission. In this case, SM merely obtains its improvement from the spatial multiplexing of the antennas. Note that there is no clear relation between increasing the antenna setting with the average percentage improvement. We showed those percentages to give the ideas of how the compared systems behave.

\begin{figure}[!t]
\centering
\includegraphics [width=0.4\textwidth]{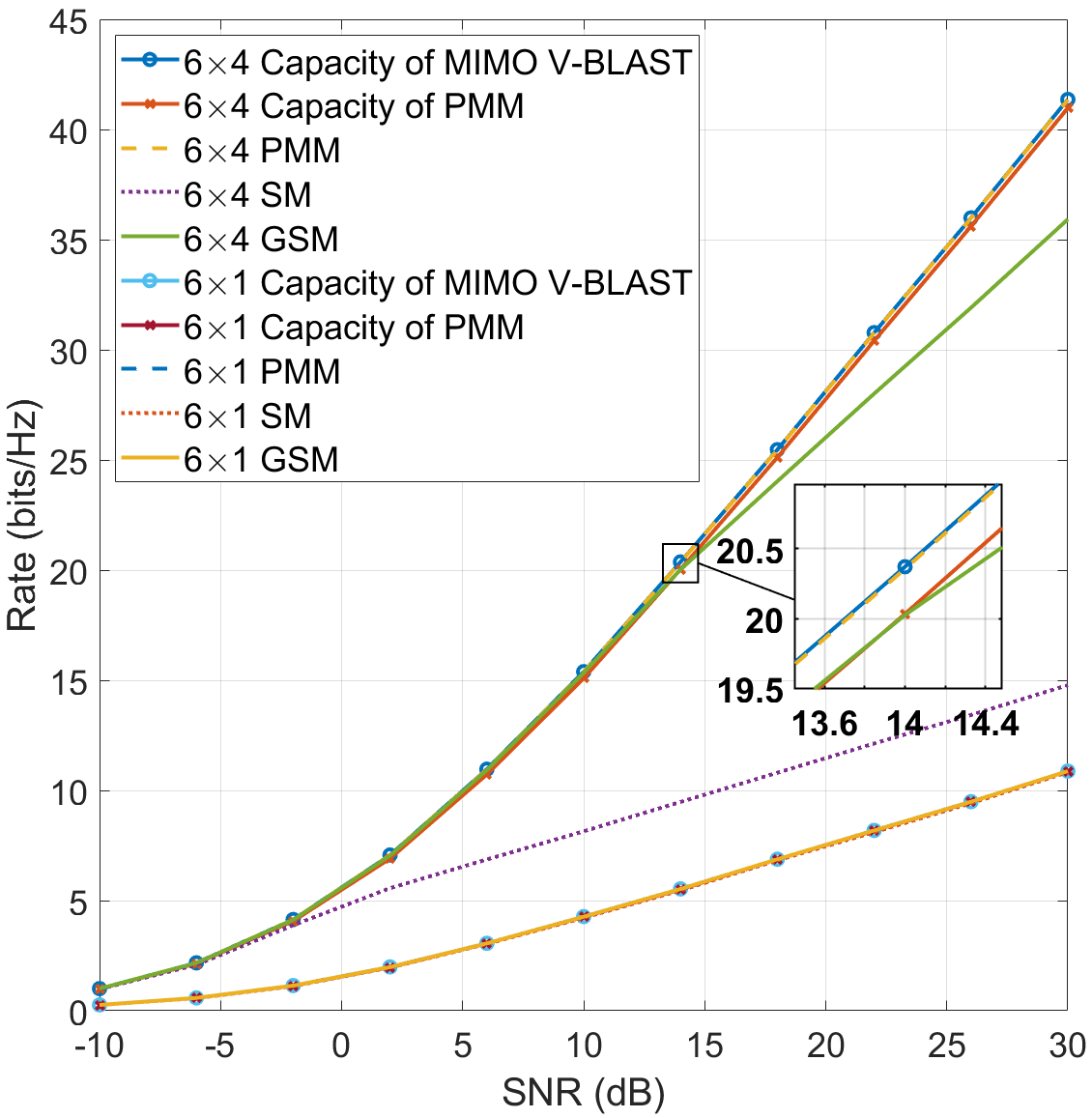} 
\caption {Comparison of capacity of MIMO V-BLAST and PMM, and the achivable rate of PMM, GSM and SM systems}
\label{rate2}
\end{figure}

In Fig. \ref{rate2}, we plot the comparison between the capacity of MIMO V-BLAST, the capacity of PMM computed using (\ref{capacity_result}), and the achievable rate of PMM, GSM, and SM under GMM distribution. The capacity of MIMO V-BLAST with Gaussian input distribution is well-known and available in many works such as in \cite{vblast}. The capacity of MIMO V-BLAST with equal power allocation can be computed as follows
\begin{equation} \label{vblast}
C_{\mathrm{MIMO}} =  \log_2 \left( \left|\mathbf{I}_{\mathrm{M}} + \frac{\rho}{N} \mathbf{HH}^{\mathrm{H}} \right| \right).
\end{equation}
Over the whole SNR range, the achievable rate of PMM has the closest performance to the capacity of MIMO V-BLAST with only $0.01$ bits/Hz gap on average. From the small box, we can also observe that the achievable rate of PMM under GMM distribution is better than its capacity. To understand these results, we must see that the overall covariance of PMM is close to equal power allocation when it is evaluated using the upper bound refinement algorithm\footnote{Typically, the lowest upper bound is obtained when all of the mixture components is merged.} which also the reason why it is very close to the capacity of MIMO V-BLAST. We can also observe that the gap between the achievable rate of PMM and GSM is closer when employing a lesser antenna. It means that PMM requires a high degree of freedom to increase its rate, just like MIMO V-BLAST. An interesting result appears when we employ MISO. All systems are almost overlapped with tiny performance differences. When we reduce the number of receive antennas to one, all systems lose their ability to exploit their spatial multiplexing gain. \revvv{It is also important to note Fig. \ref{rate2} implies that there exists at least a coding scheme for $R_0 < R_{\mathrm{PMM}}$, where $R_0$ is the desired rate. Therefore, for the area within $R_{\mathrm{SM}} < R_0 < R_{\mathrm{PMM}}$ and $R_{\mathrm{GSM}} < R_0 < R_{\mathrm{PMM}}$, a reliable communication cannot be attained by SM and GSM, respectively.}

\begin{figure}[!t]
\centering
\includegraphics [width=0.4\textwidth]{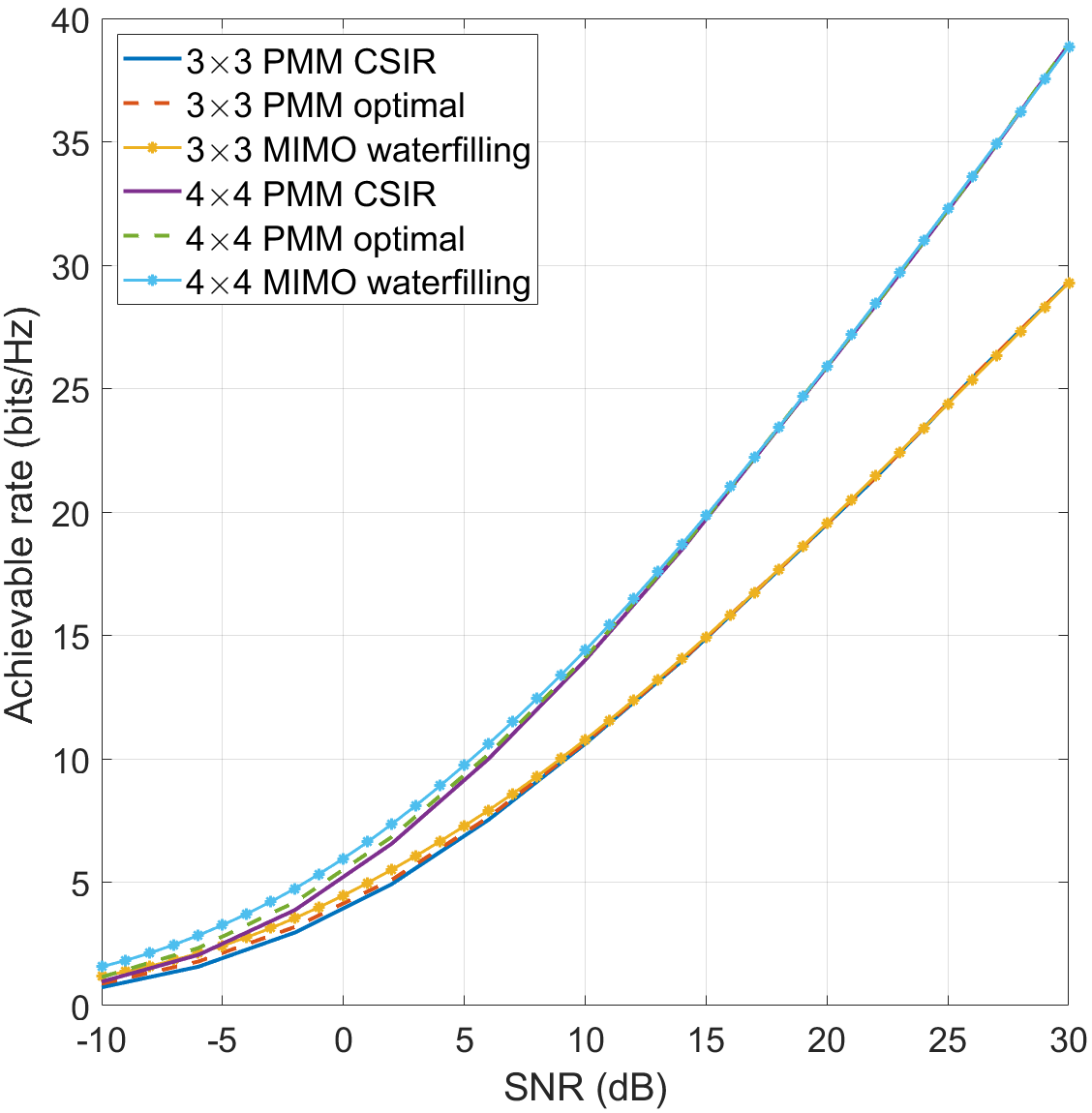} 
\caption {\rev{Comparisons of PMM CSIR, PMM optimal and MIMO waterfilling}}
\label{PMM_optim}
\end{figure}

\rev{We show the comparison between generic power allocation from TABLE \ref{pow_alloc}, optimal power allocation and MIMO waterfilling in Fig. \ref{PMM_optim}. To attain the optimal power allocation $\gamma_j^* \text{ } \forall{j}$, we assume that CSIT and CSIR are perfectly known so that we obtain a parallel channel as shown in (\ref{post_proc_csit}). We also perform the interior-point method to solve the optimization problem (\ref{optim_Rcsit2}). We can observe that the optimal achievable rate outperforms the generic power allocation over the whole SNR range. In average, the achievable rate is improved by $0.09$ bits/Hz for $3\times 3$, and $0.13$ bits/Hz for $4\times 4$. This minor improvement means that we could obtain very similar performance to the optimal achievable rate despite the absence of CSIT. Furthermore, our proposed system's performance is closely approach the performance of MIMO waterfilling despite the absence of CSIT. We can observe the effect of employing full antenna activation in PMM where we gain the benefit of full degree of freedom in the space dimension.}

\begin{figure}[!t]
\centering
\includegraphics [width=0.4\textwidth]{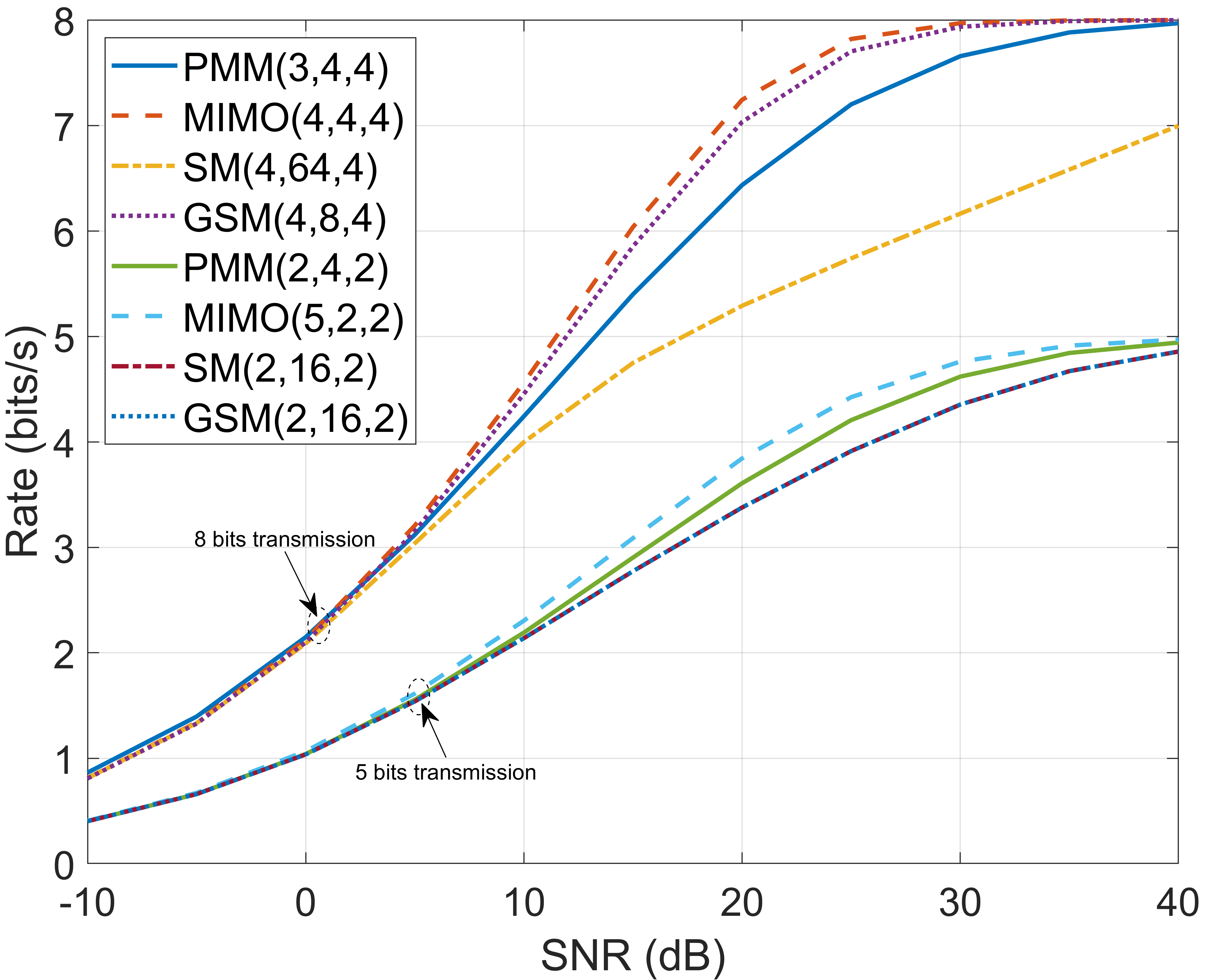} 
\caption {\revvv{Achievable rate comparisons of PMM, MIMO, GSM and SM evaluated using FCI method where X($N$,$Q$,$M$) means system X with parameter $N$, $Q$ and $M$}}
\label{AR_viaFCI}
\end{figure}

\revvv{We evaluate the achievable rate using FCI of PMM and compare it with other systems in Fig. \ref{AR_viaFCI}. The results are computed using (\ref{ARviaFCI}) where the same method is followed by all systems. From (\ref{ARviaFCI}), we can observe that each system is distinguished according to the system's symbol distances. It is worth noting that we use the same average transmit power in obtaining the results for all systems. Thus, the results are evaluated under fair circumstances. We can observe that at both $8$ and $5$ bits transmission, MIMO and SM have the best and worst performance, respectively. While PMM performs in between MIMO and SM for both transmission mode. It is important to note that PMM employs arbitrary permutation matrices with the same power allocations for all symbol combinations. The performance of PMM can still be improved by optimizing the symbol distances in (\ref{ARviaFCI}).}

\subsection{Symbol Error Rate}
\begin{figure}[!t]
\centering
\includegraphics [width=0.4\textwidth]{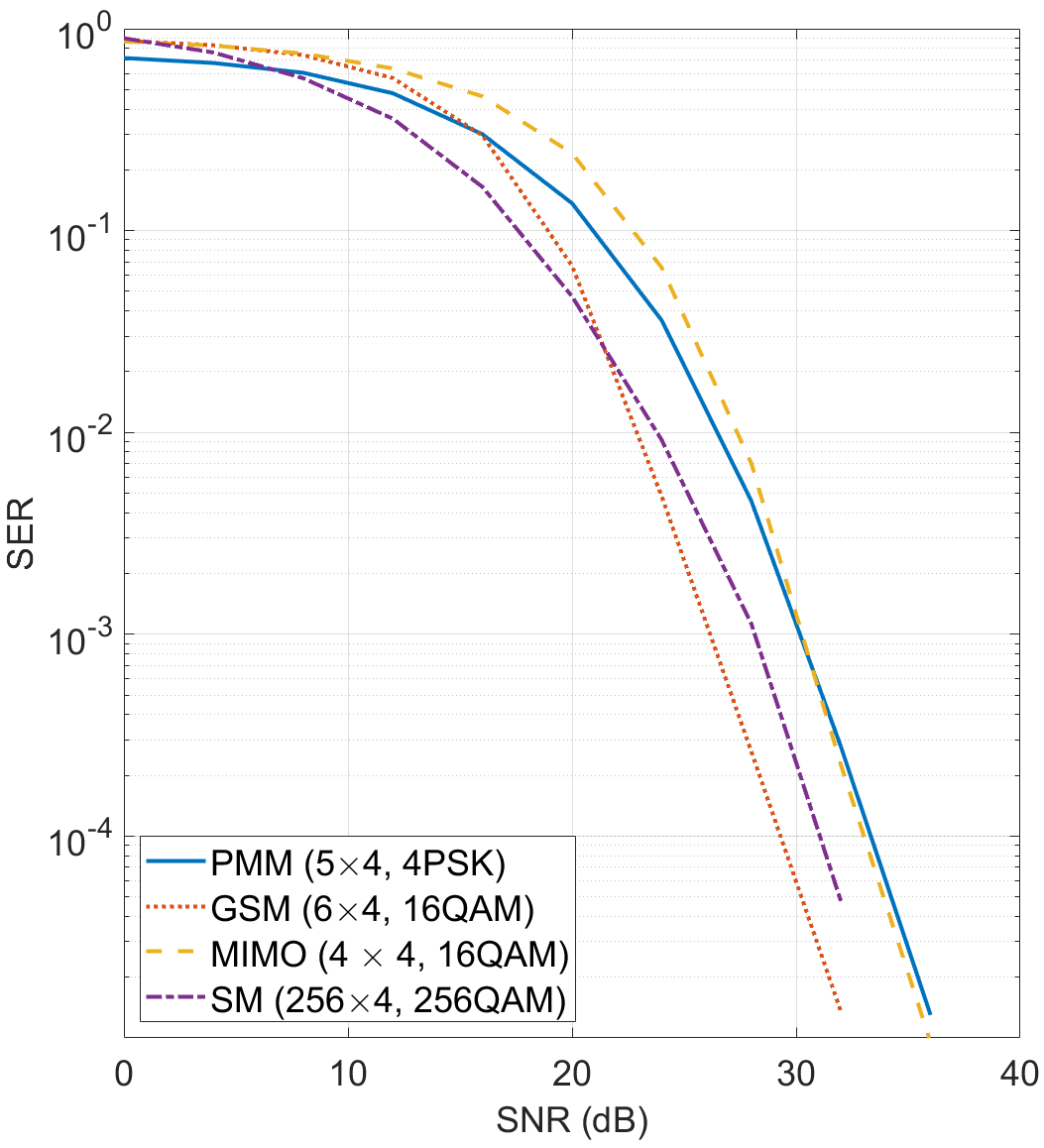} 
\caption {\rev{SER comparisons between PMM, GSM, MIMO and SM using ML detection for $16$ bits transmission}}
\label{SER_compare}
\end{figure}

\rev{We present the simulation of SER of PMM in Fig. \ref{SER_PMM} and the SER comparisons between different systems in Fig. \ref{SER_compare}. The results are evaluated by feeding $10^6$ bits to the encoder at each SNR, and we perform the Monte Carlo procedure.

In Fig. \ref{SER_PMM}, we set the antenna to be $4\times 4$ using $4$-PSK modulation. There are two different generic power allocations where the first one is defined by TABLE \ref{pow_alloc} and the second one is using $\gamma_1 = 0.45\rho$, $\gamma_2 = 0.30\rho$, $\gamma_3 = 0.15\rho$ and $\gamma_4 = 0.10\rho$. For simplicity, let us name the first power allocation (from TABLE \ref{pow_alloc}) as PA-$1$ and the second power allocation as PA-$2$. We can see in Fig. \ref{SER_PMM}, ML has a better performance compared to ZF over the whole SNR range for both power allocations. It is obvious since ML performs exhaustive search so that there is no interference left from the other streams, and the only cause of error is the noise as shown in (\ref{cor_ML}). The effect of noise can be further reduced by increasing $\rho$ to obtain higher SNR, as indicated in the figure. Unlike ML, which is a joint-detection based, ZF detects the permutation matrix and constellation symbols step-by-step. Thus, the detection of the constellation symbols depends on the correctness of the permutation matrix detection. Furthermore, the cause of error in ZF is not only the resulting noise. In fact, the resulting noise becomes colored due to the multiplication of the pseudoinverse channel matrix as shown in (\ref{post_proc_ZF}), which means the noise is increased due to this process. These are the reasons why ZF is worse than ML. An interesting result can be observed when we set different power allocations. PA-$1$ has worse performance than PA-$2$ for ML and ZF. It is straightforward to see that PA-$1$ has smaller gaps between the highest to the lowest power allocation compared to PA-$2$, where PA-$1$ has $\gamma_1 - \gamma_4 = 0.18\rho$ as the difference between the highest and the lowest power allocation, while PA-$2$ has $0.35\rho$ difference. It means that PMM results in better SER when we set more sparse power allocation for both ML and ZF detection.

\begin{figure}[!t]
\centering
\includegraphics [width=0.4\textwidth]{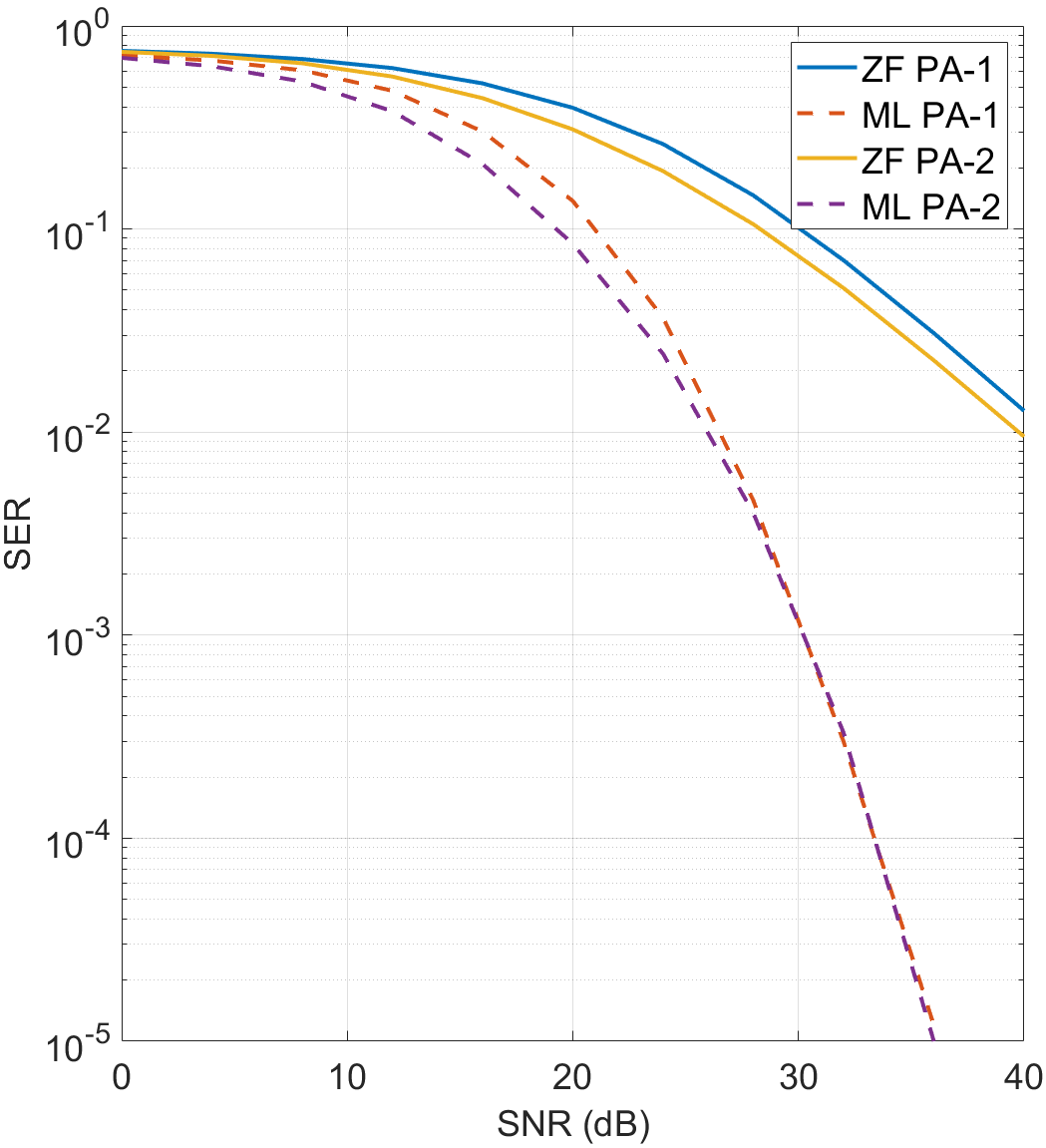} 
\caption {\rev{SER of PMM for $4\times 4$ using $4$-PSK modulation with different power allocations}}
\label{SER_PMM}
\end{figure}

Fig. \ref{SER_compare} represents the SER performance of different systems. All systems use ML detection and are set such that each system can transmit $16$ bits per transmission with the same number of receive antennas. It is worth noting that GSM activates $3$ antennas at each transmission. We can observe that the SER performance of each system varies across the SNR range. For example, PMM performs well at low SNR but declined at high SNR. An important issue to be noticed in this result is that PMM only uses 4PSK with 5 transmit antennas. While the other systems use at least 16QAM in order to maintain small antenna usage in transmitting $16$ bits. Moreover, SM needs $256$QAM and $256$ transmit antennas. This is one of the most important benefits we can gain from PMM where with far smaller constellation size or number of antennas, we can transmit the same number of bits with the other systems.}

\section{A Look on the Complexity of ML and ZF detection schemes}
In this section, we present the analysis of the complexity of ML and ZF detection schemes. The complexity can be measured with the number of floating-point operations (flops)\footnote{flop is defined as one addition, subtraction, multiplication or division of two floating-point numbers.}. We first show how to compute the complexity, and then we plot figures which represent the complexity of the corresponding detection scheme. Note that the complexity of every operation (i.e., matrix-matrix product, etc.) is computed using the standard calculations; no simplification is considered.

We can compute the complexity of ML detection by realizing that ML is based on an exhaustive search as shown in (\ref{ML}). The receive vector $\mathbf{y} \in \mathbb{C}^{M\times 1}$ is subtracted with vector, let us name it, $\hat{\mathbf{y}}_{\hat{\mathbf{P}},\hat{\mathbf{s}}} = \mathbf{H\hat{P}\Gamma\hat{s}}$ where $\mathbf{H} \in \mathbb{C}^{M \times N}$, $\hat{\mathbf{P}} \in \mathcal{P}$ of length $N \times N$ and the $k$-th element of vector $\hat{\mathbf{s}} \in \mathbb{C}^{N\times 1}$, $\hat{s}_k \in \mathcal{Q}$. Vector $\hat{\mathbf{y}}_{\hat{\mathbf{P}},\hat{\mathbf{s}}}$ must cover all possible combinations based on the inputs $\mathbf{\hat{P}}$ and $\hat{s}_k \text{ }\forall{k}$, and also the length of vector $\hat{\mathbf{s}}$ which is $N$. We know that $r$ and $Q$ are the cardinality of set $\mathcal{P}$ and $\mathcal{Q}$, respectively. Thus, there are a total of $rQ^N$ combinations to be created. To create each combination, we require $N$  flops for the product of $\mathbf{\hat{P}}$ and $\mathbf{\Gamma}$ (they are both diagonal matrices), $N^2$ flops for the product of $\mathbf{H}$ and $\hat{\mathbf{P}}\mathbf{\Gamma}$ (full matrix-diagonal matrix product) and $(2N - 1)M$ flops for the product of matrix $\mathbf{H\hat{P}\Gamma}$ and vector $\mathbf{\hat{s}}$ (full matrix-vector product). Therefore, we require a total of $N+N^2+(2N - 1)M$ flops to create a single combination for ML detection. At last, subtraction of vector $\mathbf{y}$ and vector $\hat{\mathbf{y}}_{\hat{\mathbf{P}},\hat{\mathbf{s}}}$ requires $M$ flops and the norm requires $2M-1$ flops. Finally, the total required complexity for ML is
\begin{equation} \label{ML_comp}
\bar{C}_{\mathrm{ML}} = rQ^N(N^2+N+(2N - 1)M+3M-1)
\end{equation}
flops. From (\ref{ML_comp}), we know that the complexity of ML grows by $O(rN^2Q^N)$ where $O(.)$ is the big O notation.

\begin{figure}[!t]
    \centering
    \subfloat[$M = 4, Q = 4$] {\includegraphics [width=0.45\textwidth]{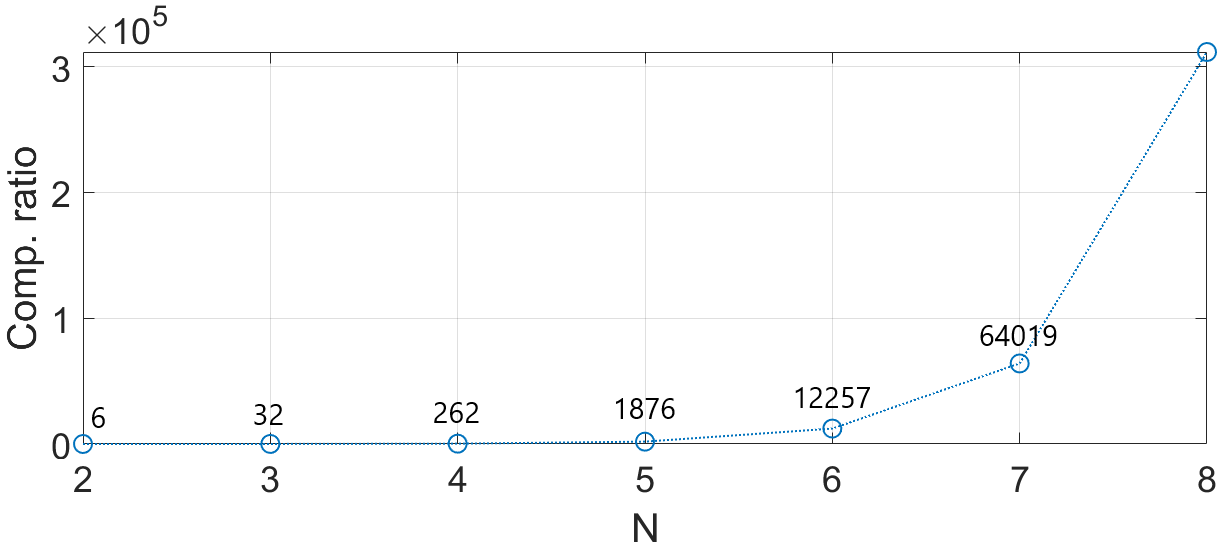}} \\
    \subfloat[$M = N = 4$] {\includegraphics [width=0.45\textwidth]{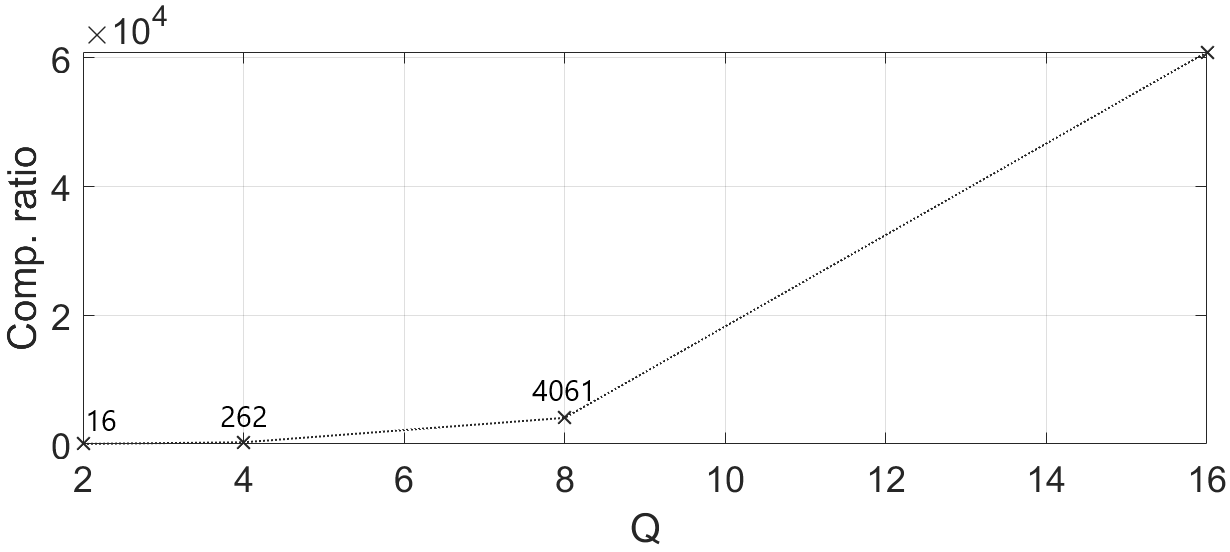}}
    \caption {Complexity ratio between ML and ZF detection}
    \label{complexity_PMM}
\end{figure}

On the other hand, ZF consists of three steps:
\begin{itemize}
\item Creation of matrix $\mathbf{H}_{\mathrm{ZF}}$ from (\ref{matrix_ZF}) which requires $N^2M+(M-1)N$ flops from $\mathbf{H}^{\mathrm{H}}\times \mathbf{H}$ (matrix-matrix product), $4M^3$ flops from $(\mathbf{H}^{\mathrm{H}}\mathbf{H})^{-1}$ (matrix inversion using the standard Gaussian elimination) and $N^2M+(M-1)N$ flops from $(\mathbf{H}^{\mathrm{H}}\mathbf{H})^{-1} \times \mathbf{H}^{\mathrm{H}}$ (matrix-matrix product). Hence, the creation of matrix $\mathbf{H}_{\mathrm{ZF}}$ requires $4M^3 + 2(N^2M+(M-1)N)$ flops in total. In the special case, when $N=M$, we can obtain $\mathbf{H}_{\mathrm{ZF}}$ by directly inverting the channel matrix $\mathbf{H}$ which will require $4M^3$ flops.
\item Detection of the permutation matrix shown by (\ref{detect_perm_M}) which requires $4M-1$ flops from diagonal matrices product, diagonal matrix-vector product, and the norm operation. This process is repeated $r$ times to cover all possible permutation matrices. Thus, this step requires a total of $r(4M-1)$ flops.
\item Detection of the constellation symbols which requires $2Q$ flops from subtracting $\tilde{y}_k$ and $\hat{s}$ and taking the norm. This process is repeated $N$ times (for all the transmitted constellation symbols). Therefore, a total of $2NQ$ is required for this step.
\end{itemize}
From all the calculations above, the total required complexity for ZF is
\begin{equation} \label{ZF_comp}
\bar{C}_{\mathrm{ZF}} = 4M^3 + 2(N^2M+(M-1)N) + r(4M-1) + 2NQ
\end{equation}
flops. From (\ref{ZF_comp}), we have that the complexity of ZF grows by $O(M^3)$.

It is important to note that the complexity difference between ML and ZF is insignificant for small antenna setting and modulation level. However, a substantial difference appears when we employ a high antenna setting or modulation level where the complexity of ML grows exponentially compared to ZF. This behavior can be observed in Fig. \ref{complexity_PMM} where we plot the complexity ratio between ML and ZF. The ratio is obtained by dividing the complexity of ML and ZF with the same parameters, $\frac{\bar{C}_{\mathrm{ML}}}{\bar{C}_{\mathrm{ZF}}}$. The ratio is increased exponentially as we increase the number of transmit antennas in Fig. \ref{complexity_PMM}(a) and the number of modulation levels in Fig. \ref{complexity_PMM}(b). For $N=8$ in Fig. \ref{complexity_PMM}(a), ML is more than $3.1 \times 10^5$ times more complex compared to ZF. This means that when we have a device with the same computational capability to detect PMM using ML and ZF, ML requires far more time than ZF.

\textit{\textbf{Remark 4:} It is typically considered that ML detection is impractical due to its computational requirement. Hence, ML is usually used to compare performance to see how good the proposed detection is in the simulation stage. In our case, we use ML to see the SER difference between ML and ZF.}

\section{Conclusion}
\rev{In this paper, we proposed a new structure of PCM named PMM, where we can modulate a set of permutation matrices to send more information bits without CSIT. A closed-form of the achievable rate of PMM was derived under a GMM distribution, and we compared the achievable rate performance with SM and GSM systems. The results suggested that PMM outperforms SM and GSM over the whole SNR range with the same parameter setting. \revvv{We also discussed the rate of PMM evaluated using FCI method where we demonstrated that with non-optimal symbol distances, PMM can approach the performance of MIMO.} An optimization problem to maximize the achievable rate was also introduced and solved using the well-known interior-point method. Furthermore, we provide the SER comparison between our proposed system with the existing systems. The result suggests that our proposed system requires the least constellation size or the number of transmit antennas in transmitting the same maximum transmit bits with a reasonable SER performance. We also proposed a low-complexity detection scheme for PMM based on ZF and demonstrated the trade-off between SER and complexity performances with ML-based detection. The results indicated that ZF detection is far less complex than ML, with a penalty in terms of the SER performance. We expect to evaluate our proposed system with more practical considerations, such as the absence of perfect CSIR and how it can work for multi-users systems as extensions to the current investigation.}



\appendices
\section{Proof of Theorem 1}
Assume that the transmit signal $\mathbf{x}$ follows a complex Gaussian distribution, we can analyze the capacity of our proposed system by having (\ref{tx_signal}) and (\ref{rx_signal}) as the input-output relation of the system. Note that since we have two sources of information, the capacity of PMM can be obtained by maximizing the mutual information between the information sources $\mathbf{P}$, $\mathbf{s}$ and the vector output $\mathbf{y}$ as follows
\begin{equation} \label{capacity}
C_{\mathrm{PMM}} = \max_{f_\mathbf{x}} I\left( \mathbf{P},\mathbf{s};\mathbf{y} \right)
\end{equation}
where $f_\mathbf{x}$ is the distribution of transmit signal $\mathbf{x}$ and
\begin{equation} \label{mut_inf_C}
I\left( \mathbf{P},\mathbf{s};\mathbf{y} \right) = H(\mathbf{y}) - H(\mathbf{y}|\mathbf{P},\mathbf{s}).
\end{equation}
We can find the second term in (\ref{mut_inf_C}) using the fact that the system is normalized, thus
\begin{equation} \label{entropy2}
H(\mathbf{y}|\mathbf{P},\mathbf{s}) = H(\mathbf{n}) = \log_2 \left|\pi e \mathbb{E}\{\mathbf{nn}^{\mathrm{H}}\} \right| =  \log_2(\left| \pi e \mathbf{I}_M \right|) 
\end{equation}
Note that (\ref{entropy2}) is independent of the input distribution $f_\mathbf{x}$, therefore the maximization in (\ref{capacity}) only relies on $H(\mathbf{y})$. Since the transmit signal $\mathbf{x} \sim \mathcal{CN}(\mathbf{0},\tilde{\mathbf{\Gamma}})$ is assumed to follow a complex Gaussian distribution, the receive signal $\mathbf{y}$ also follows complex Gaussian distribution with zero mean and covariance as shown by
\begin{equation} \label{appenA0}
\mathbb{E}\{\mathbf{yy}^{\mathrm{H}}\} = \mathbb{E}\{(\mathbf{Hx} + \mathbf{n})(\mathbf{Hx} + \mathbf{n})^\mathrm{H}\},
\end{equation}
where we know that $\mathbb{E}\{\mathbf{xx}^\mathrm{H}\} = \tilde{\mathbf{\Gamma}}$, $\mathbb{E}\{\mathbf{xn}^\mathrm{H}\} = \mathbb{E}\{\mathbf{nx}^\mathrm{H}\} = \mathbf{0}$ due to independent noise, and $\mathbb{E}\{\mathbf{nn}^\mathrm{H}\} = \mathbf{I}_\mathrm{M}$. The fact is Gaussian distribution is a differential entropy maximizer. Hence, the problem in (\ref{capacity}) is maximized by having the receive signal $\mathbf{y}$ complex Gaussian distributed. The covariance of $\mathbf{y}$ can be simplified to
\begin{equation} \label{appenA1}
\mathbb{E}\{\mathbf{yy}^{\mathrm{H}}\} = \mathbf{I}_\mathrm{M} + \mathbf{H\tilde{\Gamma}H}^\mathrm{H}.
\end{equation}
By having the covariance of $\mathbf{y}$, we can have the differential entropy of $\mathbf{y}$ as
\begin{equation} \label{appenA2}
H(\mathbf{y}) = \log_2 \left|\pi e \mathbb{E}\{\mathbf{yy}^{\mathrm{H}}\} \right| = \log_2 \left|\pi e \left( \mathbf{I}_{\mathrm{M}} +  \mathbf{H \tilde{\Gamma}H} ^{\mathrm{H}} \right) \right|.
\end{equation}
Therefore, by substituting (\ref{entropy2}) and (\ref{appenA2}) to (\ref{mut_inf_C}), we obtain the capacity of PMM under complex Gaussian distribution as shown by (\ref{capacity_result}).

\section{Proof of Theorem 2}
From (\ref{pdf_x}), we can have the conditional cumulative distribution function (cdf) of $\mathbf{x}$ given that $\mathbf{P} = \mathbf{P}_i$ as
\begin{equation} \label{cdf_x}
\begin{aligned}
F_{\mathbf{x}}^{(i)} &= p(x_1 \leq t_1,\hdots,x_N \leq t_N | \mathbf{P} = \mathbf{P}_i) \\
&= \int_{-\infty}^{t_N} \hdots \int_{-\infty}^{t_1} f_{\mathbf{x}}^{(i)} \mathrm{d}x_1 \hdots \mathrm{d}x_N
\end{aligned}
\end{equation}
where $x_k \in \mathbb{C}$ is the $k$-th element in vector $\mathbf{x}$. Using (\ref{pmf_P}), we can obtain the weighted cdf of $\mathbf{x}$ given that $\mathbf{P} = \mathbf{P}_i$ as
\begin{equation} \label{cdf_x2}
\begin{aligned}
F_{\mathbf{x}_{\alpha_i}}^{(i)} &= \alpha_i p(x_1 \leq t_1,\hdots,x_N \leq t_N | \mathbf{P} \\
&= \mathbf{P}_i) = \alpha_i \int_{-\infty}^{t_N} \hdots \int_{-\infty}^{t_1} f_{\mathbf{x}}^{(i)} \mathrm{d}x_1 \hdots \mathrm{d}x_N
\end{aligned}
\end{equation}
This leads to the marginal cdf as given by
\begin{equation} \label{marg_cdf_x}
\begin{aligned}
F_{\mathbf{x}} &= p(x_1 \leq t_1,\hdots,x_N \leq t_N ) \\
&= \sum_{i=1}^r \left( \alpha_i \int_{-\infty}^{t_N} \hdots \int_{-\infty}^{t_1} f_{\mathbf{x}}^{(i)} \mathrm{d}x_1 \hdots \mathrm{d}x_N \right).
\end{aligned}
\end{equation}
By deriving the marginal cdf in (\ref{marg_cdf_x}) and using calculus manipulation, we can have the pdf of the transmit signal $\mathbf{x}$ as shown in (\ref{pdf_gmm_x}).

\ifCLASSOPTIONcaptionsoff
  \newpage
\fi




\bibliographystyle{IEEEtran}
\bibliography{references}

%
\begin{IEEEbiography}[{\includegraphics[width=1in,height=1.25in,clip,keepaspectratio]{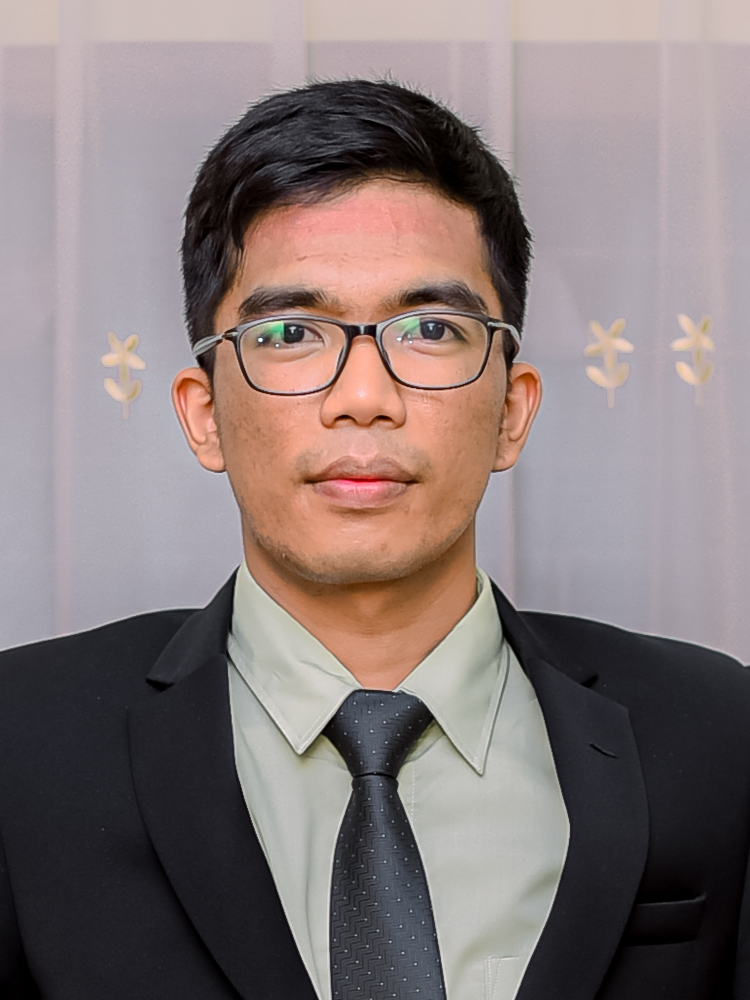}}]{Rahmat Faddli Siregar}
received his B.S. degree in telecommunication engineering in 2016 from Telkom University, Indonesia and his M.S. degree with the same field in Kumoh National Institute of Technology (KIT), South Korea, in 2018. During his M.S. study, he also joined Wireless and Emerging Network System (WENS) laboratory. Currently, he is pursuing his doctorate degree in Faculty of Information Technology and Electrical Engineering (ITEE), University of Oulu, Finland and joins Centre for Wireless Communication (CWC), one of the leading research institutes globally in the area of wireless communications. His major research interests focus on physical layer of wireless communication such as index modulation, cell-free mMIMO, waveform designs, mmWave, etc.
\end{IEEEbiography}


\begin{IEEEbiography}[{\includegraphics[width=1in,height=1.25in,clip,keepaspectratio]{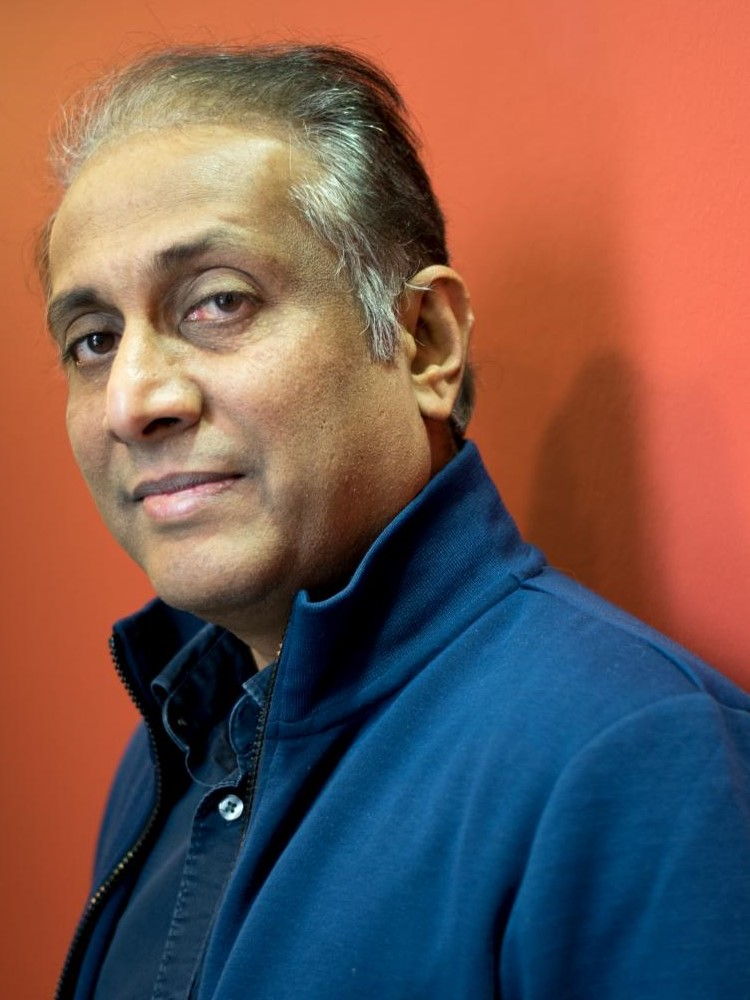}}]{Nandana Rajatheva}
(Senior Member, IEEE) received the B.Sc. degree (Hons.) in electronics and telecommunication engineering from the University of Moratuwa, Sri Lanka, in 1987, and the M.Sc. and Ph.D. degrees from the University of Manitoba, Winnipeg, MB, Canada, in 1991 and 1995, respectively. He is currently a Professor with the Centre for Wireless Communications, University of Oulu, Finland. During his graduate studies, he was a Canadian Commonwealth Scholar in Manitoba. From 1995 to 2010, he was a Professor/an Associate Professor with the University of Moratuwa and the Asian Institute of Technology, Thailand. He is currently leading the AI-driven air interface design task in Hexa-X EU Project. He has coauthored more than 200 refereed papers published in journals and in conference proceedings. His research interests include physical layer in beyond 5G, machine learning for PHY and MAC, integrated sensing and communications, and channel coding.
\end{IEEEbiography}


\begin{IEEEbiography}[{\includegraphics[width=1in,height=1.25in,clip,keepaspectratio]{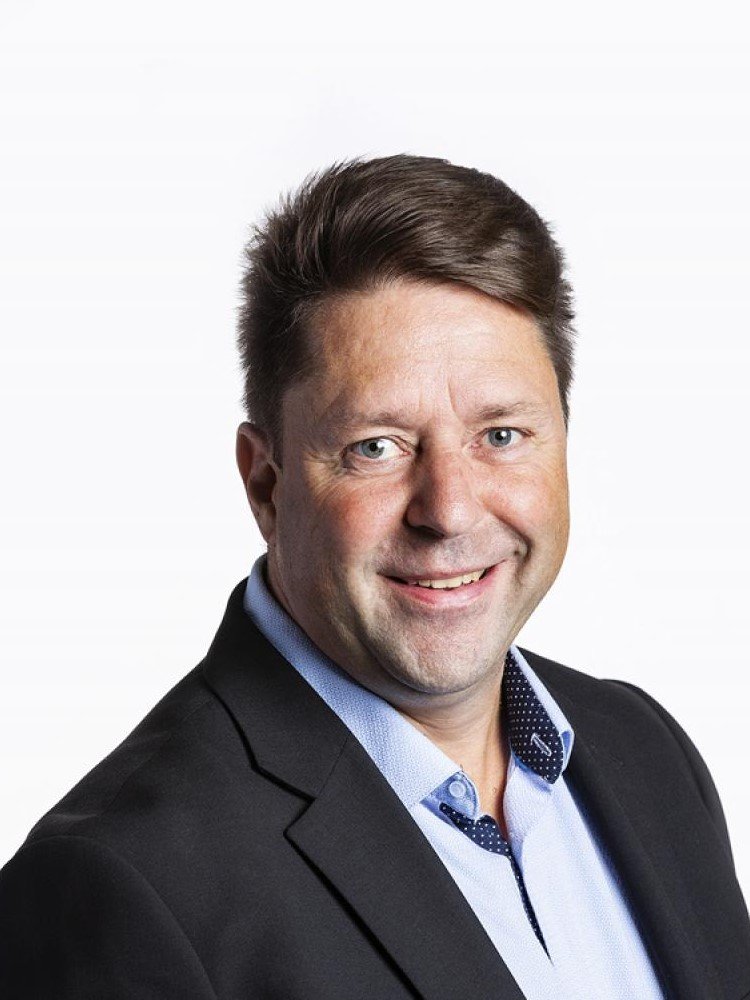}}]{Matti Latva-Aho}
(Senior Member, IEEE) received the M.Sc., Lic.Tech., and Dr.Tech. (Hons.) degrees in electrical engineering from the University of Oulu, Finland, in 1992, 1996, and 1998, respectively. From 1992 to 1993, he was a Research Engineer at Nokia Mobile Phones, Oulu, Finland, after that he joined the Centre for Wireless Communications (CWC), University of Oulu. He was the Director of CWC, from 1998 to 2006, and the Head of the Department for Communication Engineering, until August 2014. Currently, he serves as an Academy of Finland Professor and is the Director for National 6G Flagship Program. He is also a Global Fellow with Tokyo University. His research interests include mobile broadband communication systems and currently his group focuses on 6G systems research. He has published over 500 journals or conference papers in the field of wireless communications. In 2015, he received the Nokia Foundation Award for his achievements in mobile communications research.
\end{IEEEbiography}

\end{document}